\documentclass[aps,english,preprint,nofootinbib,preprintnumbers]{revtex4-2}
\usepackage{mathrsfs}
\usepackage{hyperref}
\hypersetup{
	colorlinks=true,
	linkcolor=blue,
	filecolor=magenta,      
	urlcolor=cyan,
	pdfpagemode=FullScreen,
}
\usepackage{graphicx}
\usepackage{amsmath, amssymb,bbm}
\usepackage{babel}
\usepackage{color}
\usepackage{slashed}
\usepackage[T1]{fontenc}
\usepackage{blindtext}
\usepackage{titlesec}
\usepackage{empheq}
\usepackage{multirow}
\usepackage{slashbox}
\def\beqn{\begin{eqnarray}}
	\def\eeqn{\end{eqnarray}}
\def\barr{\begin{array}}
	\def\earr{\end{array}}
\def\btab{\begin{tabular}}
	\def\etab{\end{tabular}}
\def\bite{\begin{itemize}}
	\def\eite{\end{itemize}}
\def\bcen{\begin{center}}
	\def\ecen{\end{center}}

\newcommand{\pe}{\mathbf{p_e}}
\newcommand{\pnup}{\mathbf{p_\nu'}}

\begin{document}
	
\title{Most general neutron decay correlations: Standard Model recoil and radiative corrections}

\author{Chien-Yeah Seng$^{1,2}$}

\affiliation{$^{1}$Facility for Rare Isotope Beams, Michigan State University, East Lansing, MI 48824, USA}
\affiliation{$^{2}$Department of Physics, University of Washington,
Seattle, WA 98195-1560, USA}

\date{\today}

\begin{abstract}
		
To continue from our previous work \textit{Phys.Rev.D\textbf{109}(2024),073007}, we derive the full Standard Model prediction of the most general free neutron differential decay rate with all massive particles (neutron, proton and electron) polarized, including the $\mathcal{O}(1/m_N)$ recoil corrections and $\mathcal{O}(\alpha/\pi)$ radiative corrections. For the latter we adopt the newly-developed pseudo-neutrino formalism which is compatible to realistic experimental setups, in which neutrinos and photons are not detected. We also provide readily-executable \textit{Mathematica} notebooks to evaluate these corrections.  
		
\end{abstract}

\maketitle
\newpage
	
	
\section{Introduction}

This paper is a direct sequel to our previous work, Ref.\cite{Seng:2024fvi}. In that paper, we studied for the first time consequences to the free neutron decay assuming that the polarization of the outcoming proton could be measured. This gives rise to a very rich decay correlation structure which can be used as a powerful tool not only to probe new physics beyond the Standard Model (BSM), but also to test the consistency of the effective field theory (EFT) description of BSM physics (which assumes new degrees of freedoms (DOFs) are heavy) and to search for signals of light new DOFs.
This effort was partially motivated by the recent discrepancy in the determination of the axial-to-vector ratio $\lambda$ from the electron-neutrino correlation $a$~\cite{Beck:2019xye,Beck:2023hnt} and the beta asymmetry parameter $A$~\cite{Markisch:2018ndu}, which is difficult to be explained within the EFT framework.

To make full use of this new formalism, one requires a precise Standard Model (SM) prediction of the new correlation coefficients in order to isolate the small BSM effects from experimental measurements. In Ref.\cite{Seng:2024fvi} we studied only the tree-level SM contributions, accompanied by the Fermi function~\cite{Fermi:1934hr} and the virtual radiative corrections. In this paper we complete the task by including all the SM higher-order corrections up to $10^{-4}$, which cover the full $\mathcal{O}(1/m_N)$ recoil corrections and the $\mathcal{O}(\alpha/\pi)$ radiative corrections. While the former is straightforward, the latter is more complicated as its depends on the actual experimental setup. In particular, it was recently pointed out~\cite{Gluck:2022ogz} that the ``conventional'' treatment of the so-called outer radiative corrections~\cite{Sirlin:1967zza,Shann:1971fz,Garcia:1978bq} is incompatible to actual experiments as it depends on the neutrino momentum $\vec{p}_\nu$ (the ``neutrino'' formalism) which is never actually measured, and cannot be deduced directly from the electron and proton momenta when an extra (undetected) photon is emitted. To circumvent this problem, the differential decay rate must be expressed in terms of fully-measurable quantities; possible choices are $\{\vec{p}_e,E_p\}$ (the ``recoil'' formalism)~\cite{Toth:1984er,Gluck:1992tg,Gluck:1989sf,Gluck:1992qy,Gluck:1998ogp,Gluck:1997km,Gluck:1994sw,Gluck:2022ogz} and $\{\vec{p}_e,\Omega_\nu'\}$ (the ``pseudo-neutrino'' formalism)~\cite{Seng:2023ynd}, where $\Omega_\nu'$ is the solid angle of the ``pseudo-neutrino'' momentum $\vec{p}_\nu'\equiv -\vec{p}_e-\vec{p}_p$. The applicability of the first method is more restrictive because half of the angular observables are integrated out, which make it difficult to describe many spin-dependent correlations (e.g. the neutrino asymmetry parameter $B$). The pseudo-neutrino formalism, on the other hand, is capable to describe all correlations of interest and thus will be adopted in this work. 

The content of this work is arranged as follows. In Sec.\ref{sec:recoil}, \ref{sec:RC} we lay out the theory framework for the $\mathcal{O}(1/m_N)$ recoil corrections and $\mathcal{O}(\alpha/\pi)$ radiative corrections, respectively; the full SM expression including such corrections are presented in Sec.\ref{sec:full}. In Sec.\ref{sec:check} we compare our results to all existing results of recoil and radiative corrections (to the best of our knowledge) that do not involve the proton polarization. A brief summary is provided in Sec.\ref{sec:summary}. Some technical details in this work, as well as some basic instructions to utilize the supplemented \textit{Mathematica} notebooks, are given in the Appendices.

\section{\label{sec:recoil}Framework for recoil corrections}

In this section we describe the theory foundation for the 3-body final state contribution ($n(p_n)\rightarrow p(p_p)+e(p_e)+\bar{\nu}(p_\nu)$) to the neutron differential decay rate, that allows us to incorporate the $\mathcal{O}(1/m_N)$ recoil corrections. Our derivation follows closely to that in Ref.\cite{Ando:2004rk}. We start from the 3-body decay rate formula:
\begin{equation}
	\Gamma_3=\frac{1}{2m_n}\int\frac{d^3p_p}{(2\pi)^3 2E_p}\frac{d^3p_e}{(2\pi)^3 2E_e}\frac{d^3p_\nu}{(2\pi)^3 2E_\nu}(2\pi)^4\delta^{(4)}(p_n-p_p-p_e-p_\nu)|\mathcal{M}_3|^2\equiv \int d\Pi_3|\mathcal{M}_3|^2~,
\end{equation}
where $\mathcal{M}_3$ is the 3-body decay amplitude.
Using the spatial delta function to integrate out $\vec{p}_\nu$, we can simplify the 3-body phase space as:
\begin{eqnarray}
	\int d\Pi_3&=&\frac{1}{16(2\pi)^5m_n}\int d\Omega_e d\Omega_\nu \int_{m_e}^{E_m}dE_e\pe \int dE_\nu\frac{E_\nu}{m_n-E_e+\pe c}\times \nonumber\\
	&&\delta\left(E_\nu-\frac{E_m-E_e}{1-\frac{E_e-\pe c}{m_n}}\right)~,\label{eq:Pi3}
\end{eqnarray}
where $E_m\equiv (m_n^2-m_p^2+m_e^2)/(2m_n)$ is the \textit{exact} electron end-point energy, $\pe\equiv |\vec{p}_e|$, and  $c\equiv \hat{p}_e\cdot\hat{p}_\nu$ is the cosine of the angle between the electron and neutrino momenta. Notice that the energy delta function does not impose any constraint to the solid angles $\Omega_e$, $\Omega_\nu$, since $E_\nu$ remains positive for all values of $c$. 

The recoil corrections come from both the squared amplitude and the phase space, and let us start with the former. The SM 3-body amplitude reads:
\begin{equation}
	\mathcal{M}_3=-\frac{G_V}{\sqrt{2}}L_\mu H^\mu~.
\end{equation}
Let us explain the notations. First, $G_V=G_F V_{ud} g_V$, where $G_F=1.1663788(6)\times 10^{-5}$~GeV$^{-2}$ is the Fermi coupling constant measured from muon decay~\cite{MuLan:2012sih}, $V_{ud}$ is the upper left component of the Cabibbo-Kobayashi-Maskawa matrix~\cite{Cabibbo:1963yz,Kobayashi:1973fv}, and $g_V$ is the neutron vector coupling constant (more explanations later); 
$L_\mu=\bar{u}_e\gamma_\mu(1-\gamma_5)v_\nu$ and $H_\mu=\bar{u}_p\Gamma_\mu u_n$ are the matrix elements of the leptonic and hadronic charged weak current respectively, with the nucleon vertex function defined as:
\begin{equation}
	\Gamma^\mu(p_p,p_n)\equiv\gamma^\mu(1+\lambda\gamma_5)-\frac{i}{2m_N}(\mu_V-1)\sigma^{\mu\nu}(p_n-p_p)_\nu-\frac{2m_N\lambda}{m_\pi^2}(p_n-p_p)^\mu\gamma_5~,\label{eq:Gammamu}
\end{equation}
where $\lambda\equiv g_A/g_V<0$ is the axial-to-vector coupling ratio (more explanations later), $\mu_V=\kappa_V+1\equiv \mu_p-\mu_n\approx 4.7059$ is the weak magnetic moment, $m_N\equiv (m_n+m_p)/2$ is the averaged nucleon mass, and $m_\pi$ is the pion mass. The pseudoscalar coupling is related to the axial coupling through the partially-conserved axial current (PCAC) relation. Notice that we have dropped the momentum-dependence of the nucleon form factors in the expression above, because its effect scales as $E_e^2/\Lambda^2<10^{-4}$, where $\Lambda$ is the relevant hadronic mass scale in the form factors. {\color{black}Also, we neglect the so-called ``second-class currents'', namely the induced-scalar and induced-tensor form factors, as they are suppressed simultaneously by recoil and isospin symmetry breaking.}
Another important point is that, we have defined the vector and axial couplings above to include the ``inner'' radiative corrections:
\begin{equation}
	g_V^2=\mathring{g}_V^2(1+\Delta_R^V)~,~g_A^2=\mathring{g}_A^2(1+\Delta_R^A)~,
\end{equation}
where $\mathring{g}_{V,A}$ are the pure Quantum Chromodynamics (QCD)-induced vector and axial couplings; in particular, $\mathring{g}_V=1$ barring a possibly-relevant strong isospin symmetry breaking correction that can be studied using lattice QCD~\cite{Seng:2023jby}. Tremendous progress is observed in recent years to the inner corrections: see, e.g. Refs.\cite{Seng:2018yzq,Seng:2018qru,Czarnecki:2019mwq,Seng:2020wjq,Hayen:2020cxh,Cirigliano:2023fnz} for $\Delta_R^V$ and Refs.\cite{Hayen:2020cxh,Gorchtein:2021fce,Cirigliano:2022hob,Seng:2024ker} for $\Delta_R^A$.

An efficient way to keep track of the recoil corrections from the squared amplitude is to factor out $4m_n m_p$ from $|\mathcal{M}_3|$ and define a ``quantum mechanical'' squared amplitude {\color{black}(which corresponds to properly-normalized external states, $\langle \phi|\phi\rangle =1$)}:
\begin{equation}
	|\mathcal{M}_3|^2_\text{QM}\equiv \frac{1}{4m_n m_p}|\mathcal{M}_3|^2
	=\frac{G_V^2}{2}L_{\mu\nu}H^{\mu\nu}~,
\end{equation}
where\footnote{The $H_{\mu\nu}$ here differs from that in Ref.\cite{Seng:2023ynd} by a factor $1/(4m_nm_p)$.}
\begin{eqnarray}
	L_{\mu\nu}&\equiv&\sum_{s_\nu}L_\mu L_\nu^*=\text{Tr}\left[\Sigma_e(\slashed{p}_e+m_e)\gamma_\mu(1-\gamma_5)\slashed{p}_\nu\gamma_\nu(1-\gamma_5)\right]\nonumber\\
	H_{\mu\nu}&\equiv&\frac{1}{4m_n m_p}H_\mu H_\nu^*=\frac{1}{4m_nm_p}\text{Tr}\left[\Sigma_p(\slashed{p}_p+m_p)\Gamma_\mu \Sigma_n(\slashed{p}_n+m_n)\overline{\Gamma}_\nu\right]~,\label{eq:LH}
\end{eqnarray}
with $\overline{\Gamma}_\nu\equiv \gamma^0\Gamma_\nu^\dagger\gamma^0$. In this work we assume $n$, $p$ and $e$ are all polarized, which introduce three spin projection operators:
\begin{equation}
	\Sigma_\phi=\frac{1}{2}(1+\gamma_5\slashed{s}_\phi)~,~\phi=n,p,e~,
\end{equation}
with the spin vector $s_\phi^\mu$ satisfying $s_\phi\cdot p_\phi=0$, $s_\phi^2=0$. In the neutron's rest frame they take the following expressions:
\begin{eqnarray}
	s_n^\mu&=&(0,\hat{s}_n)\nonumber\\
	s_p^\mu&=&\left(\frac{\vec{p}_p\cdot\hat{s}_p}{m_p},\hat{s}_p+\frac{\vec{p}_p\cdot\hat{s}_p\:\vec{p}_p}{m_p(E_p+m_p)}\right)\nonumber\\
	&\approx&\left(\frac{\vec{p}_p\cdot\hat{s}_p}{m_N},\hat{s}_p\right)\nonumber\\
	s_e^\mu&=&\left(\frac{\vec{p}_e\cdot\vec{\sigma}}{m_e},\vec{\sigma}+\frac{\vec{p}_e\cdot\vec{\sigma}\:\vec{p}_e}{m_e(E_e+m_e)}\right)~,
\end{eqnarray}
with $\hat{s}_n$, $\hat{s}_p$ and $\vec{\sigma}$ the unit polarization vector of $n$, $p$ and $e$ in their respective rest frame.

One may now expand the squared amplitude as
\begin{equation}
	|\mathcal{M}_3|^2_\text{QM}=E_e E_\nu\left\{F^{(0)}(\vec{p}_e,\Omega_\nu)+\frac{1}{m_N}F^{(1)}(\vec{p}_e,E_\nu,\Omega_\nu)+...\right\}~.\label{eq:M3expand}
\end{equation}
$F^{(0)}$ is the result in the non-recoil limit, while $F^{(1)}$ contains the leading recoil correction to the squared amplitude. The expansion is most easily doable by first expressing the proton momentum as:
\begin{equation}
	p_p=p_n-p_e-p_\nu=(m_n-E_e-E_\nu,-\vec{p}_e-\vec{p}_\nu)~,
\end{equation}
and writing $m_n=m_N+\delta m/2$, $m_p=m_N-\delta m/2$, with $\delta m\equiv m_n-m_p$. After that, we rescale the hadron masses as\footnote{Although $m_\pi\ll m_N$, we have checked that the contribution from the pseudoscalar form factor scales as $E_e^2/m_\pi^2<10^{-4}$, which can be dropped given our precision goal.}:
\begin{equation}
	m_N\rightarrow m_N/\epsilon~,~m_\pi\rightarrow m_\pi/\epsilon~,\label{eq:epsilon}
\end{equation}
and expand $|\mathcal{M}_3|_\text{QM}^2$ to $\mathcal{O}(\epsilon)$ to get Eq.\eqref{eq:M3expand}.
It is useful to note that $F^{(0)}$ depend only on $\vec{p}_e$, $\Omega_\nu$ and not on $E_\nu$. This is because in the non-recoil limit the neutrino momentum appears linearly (and only linearly) through the lepton tensor as $\slashed{p}_\nu$. Since $p_\nu^\mu=E_\nu(1,\hat{p}_\nu)$, so after scaling out $E_\nu$ the rest must be $E_\nu$-independent. 

We plug Eq.\eqref{eq:M3expand} back into Eq.\eqref{eq:Pi3} and perform the following expansion~\cite{Ando:2004rk}:
\begin{equation}
	\frac{m_p E_\nu^2}{m_n-E_e+\pe c}\approx(E_m-E_e)^2\left\{1+\frac{3E_e-E_m-3\pe c}{m_N}\right\}
\end{equation}
which encodes the leading recoil corrections from the phase space. That gives the following master formula
\begin{eqnarray}
	\Gamma_3&=&\frac{1}{128\pi^5}\int d\Omega_e d\Omega_\nu\int_{m_e}^{E_m} dE_e\pe E_e(E_m-E_e)^2\times\nonumber\\
	&&\left\{F^{(0)}(\vec{p}_e,\Omega_\nu)+\frac{(3E_e-E_m-3\pe c)F^{(0)}(\vec{p}_e,\Omega_\nu)+F^{(1)}(\vec{p}_e,E_m-E_e,\Omega_\nu)}{m_N}+...\right\}\label{eq:recoilmaster}
\end{eqnarray}
that accounts for the first-order recoil effects in the 3-body decay from both the phase space and the squared amplitude. Notice that we have replaced $E_\nu\rightarrow E_m-E_e$ in $F^{(1)}$ since this term is already suppressed by $1/m_N$. 

We include, as a supplementary material, a \textit{Mathematica} notebook \textbf{recoil.nb} to demonstrate explicitly all the steps above. 

\section{\label{sec:RC}Framework for radiative corrections}

Next we study the $\mathcal{O}(\alpha/\pi)$ radiative corrections, which include both one-loop and bremsstrahlung corrections; the two have to be added to ensure the infrared-finiteness of the final result. The former has already been studied  in Ref.\cite{Seng:2024fvi} and here we only need to compute the bremsstrahlung contribution which concerns the 4-body decay $n(p_n)\rightarrow p(p_p)+e(p_e)+\bar{\nu}(p_\nu)+\gamma(k)$.

We start from 4-body decay rate formula:
\begin{eqnarray}
	\Gamma_4&=&\frac{1}{2m_n}\int\frac{d^3p_p}{(2\pi)^3 2E_p}\frac{d^3p_e}{(2\pi)^3 2E_e}\frac{d^3p_\nu}{(2\pi)^3 2E_\nu}\frac{d^3k}{(2\pi)^3 2E_k}(2\pi)^4\delta^{(4)}(p_n-p_p-p_e-p_\nu-k)|\mathcal{M}_4|^2\nonumber\\
	&\equiv& \int d\Pi_4 |\mathcal{M}_4|^2~.
\end{eqnarray}
As advertised in the Introduction, we adopt the pseudo-neutrino formalism~\cite{Seng:2023ynd} by defining an experimentally-measurable pseudo-neutrino momentum: 
\begin{equation}
p_\nu'\equiv p_n-p_p-p_e~\equiv (E_\nu',\vec{p}_\nu')~.
\end{equation}
In 3-body decay, $p_\nu'$ and $p_\nu$ are equivalent so nothing in Sec.\ref{sec:recoil} has to be changed. On the other hand, the 4-body phase space takes the following form in the non-recoil limit:
\begin{eqnarray}
	\int d\Pi_4&\approx&\frac{1}{512\pi^6 m_n m_p}\int d\Omega_e d\Omega_\nu'\int_{m_e}^{E_m}dE_e\pe \int_0^{E_\nu'}d\pnup \pnup^2 \int\frac{d^3k}{(2\pi)^3 2E_k}\frac{d^3p_\nu}{(2\pi)^32 E_\nu}\nonumber\\
	&&\times (2\pi)^4\delta^{(4)}(p_\nu'-p_\nu-k)~,
\end{eqnarray}
where $\pnup\equiv|\vec{p}_\nu'|$, and notice that $\pnup\neq E_\nu'$ in general.
Rigorously speaking, the upper limit of $E_e$ should be taken as the ``zeroth-order'' electron end-point energy $E_m^0\equiv m_n-m_p$ in the non-recoil limit, but since the 4-body contribution itself is already $\mathcal{O}(\alpha/\pi)$-suppressed, one may replace $E_m^0\rightarrow E_m$ to be consistent with the 3-body formula, which results only in a $\mathcal{O}(\alpha/\pi\times E_e/m_N)$ correction that can be neglected. With the same reason, one simply takes $E_\nu'\approx E_m-E_e$ everywhere in the $\mathcal{O}(\alpha/\pi)$ contribution.

The 4-body amplitude, after neglecting recoil corrections, takes the following form:
\begin{eqnarray}
	\mathcal{M}_4&\approx& -\frac{G_V e}{\sqrt{2}}\left(\frac{p\cdot\varepsilon^*}{p\cdot k}-\frac{p_e\cdot \varepsilon^*}{p_e\cdot k}\right)H_\mu L^\mu+\frac{G_Ve}{\sqrt{2}}\left(\frac{k^\mu\varepsilon^{*\nu}-k^\nu\varepsilon^{*\mu}-i\epsilon^{\mu\nu\alpha\beta}k_\alpha \varepsilon_\beta^*}{2p_e\cdot k}\right)H_\mu L_\nu\nonumber\\
	&\equiv & \mathcal{M}_{4\text{I}}+\mathcal{M}_{4\text{II}}~,
\end{eqnarray}
where we may drop the weak magnetism and pseudoscalar form factors in $H_\mu$ as they are recoil-suppressed. Also, we take $p_n,p_p\rightarrow p$ in the 4-body amplitude with $p^2=m_N^2$.   
Following Ref.\cite{Seng:2023ynd}, we split the squared amplitude as:
\begin{equation}
	|\mathcal{M}_4|_\text{QM}^2\equiv \frac{1}{4m_N^2}|\mathcal{M}_4|^2=|\mathcal{M}_{4\text{I}}|_\text{QM}^2+2\mathfrak{Re}\left\{\mathcal{M}_{4\text{I}}\mathcal{M}_{4\text{II}}^*\right\}_\text{QM}+|\mathcal{M}_{4\text{II}}|_\text{QM}^2~,
\end{equation}
where
\begin{eqnarray}
	|\mathcal{M}_\text{4I}|_\text{QM}^2&=&-\frac{G_V^2e^2}{2}\left(\frac{p}{p\cdot k}-\frac{p_e}{p_e\cdot k}\right)^2H_{\mu\nu}L^{\mu\nu}\equiv 	|\mathcal{M}_\text{4I}|_{\text{QM},a}^2+	|\mathcal{M}_\text{4I}|_{\text{QM},b}^2 \nonumber\\
	2\mathfrak{Re}\left\{\mathcal{M}_{4\text{I}}\mathcal{M}_{4\text{II}}^*\right\}_\text{QM}&=&G_V^2e^2\mathfrak{Re}\left\{\left(\frac{p}{p\cdot k}-\frac{p_e}{p_e\cdot k}\right)_\alpha\left(\frac{k^\rho g^{\alpha\sigma}-k^\sigma g^{\rho\alpha}+i\epsilon^{\rho\sigma\gamma \alpha}k_\gamma}{2p_e\cdot k}\right)H_{\mu\rho}L^\mu_{\:\:\sigma} \right\}\nonumber\\
	|\mathcal{M}_\text{4II}|_\text{QM}^2&=&-\frac{G_V^2e^2}{8(p_e\cdot k)^2}\left\{k^\mu g^{\nu\alpha}-k^\nu g^{\mu\alpha}-i\epsilon^{\mu\nu\gamma\alpha}k_\gamma\right\}\times\nonumber\\
	&&\left\{k^\rho g^\sigma_{\:\:\alpha}-k^\sigma g^\rho_{\:\:\alpha}+i\epsilon^{\rho\sigma\delta}_{~~~\:\alpha}k_\delta\right\}H_{\mu\rho}L_{\nu\sigma}~.
\end{eqnarray}
Notice that, in the first equation we have substituted $p_\nu=p_\nu'-k$ in $L_{\mu\nu}$, and the ``$a$'' and ``$b$'' term correspond to taking the $p_\nu'$ and $-k$ piece, respectively. 

Among various terms in the bremsstrahlung contribution, only $|\mathcal{M}_{4\text{I}}|^2_{\text{QM},a}$ is infrared-divergent upon integrating over $p_\nu$, $k$ and $\pnup$, but such integrals can be analytically performed. It combines with the virtual radiative corrections to yield an infrared-finite result (see Ref.\cite{Seng:2023ynd} for details). For the remaining, ``regular'' piece: 
\begin{equation}
	|\mathcal{M}_4|_\text{QM,reg}^2\equiv |\mathcal{M}_\text{4I}|_{\text{QM},b}^2+2\mathfrak{Re}\left\{\mathcal{M}_{4\text{I}}\mathcal{M}_{4\text{II}}^*\right\}_\text{QM}+|\mathcal{M}_\text{4II}|_\text{QM}^2~,\label{eq:M4reg}
\end{equation}
the $p_\nu$- and $k$-integral can be done analytically (see Appendix~\ref{sec:scalar}, \ref{sec:tensor} for details), and the remaining one-fold integration over $\pnup$ is most conveniently carried out numerically. The full steps to obtain the contribution from $|\mathcal{M}_4|_{\text{QM,reg}}^2$ are provided in a second supplemented \textit{Mathematica} notebook, \textbf{bremreg.nb}.

\section{\label{sec:full}Full expression}

\begin{table}
	\begin{centering}
		\begin{tabular}{|c|c|c|c|}
			\hline 
			$\mathcal{C}$ & $\mathcal{C}_{0}$ & $\delta_{\text{an},1}^{\mathcal{C}}$ & $\delta_{\text{an},2}^{\mathcal{C}}$\tabularnewline
			\hline 
			\hline 
			$a$ & $\frac{1-\lambda^{2}}{1+3\lambda^{2}}$ & $\frac{2(1-\beta^{2})}{\beta}\tanh^{-1}\beta+4(1-\ln4)\left(\frac{1}{\beta}\tanh^{-1}\beta-1\right)$ & 0\tabularnewline
			\hline 
			$A$ & $\frac{-2\lambda(\lambda+1)}{1+3\lambda^{2}}$ & $\frac{2(1-\beta^{2})}{\beta}\tanh^{-1}\beta$ & 0\tabularnewline
			\hline 
			$B$ & $\frac{2\lambda(\lambda-1)}{1+3\lambda^{2}}$ & $4(1-\ln4)\left(\frac{1}{\beta}\tanh^{-1}\beta-1\right)$ & 0\tabularnewline
			\hline 
			$G$ & $-1$ & $\frac{2(1-\beta^{2})}{\beta}\tanh^{-1}\beta$ & 0\tabularnewline
			\hline 
			$H$ & $\frac{\lambda^{2}-1}{1+3\lambda^{2}}\frac{m_{e}}{E_{e}}$ & $-2\beta\tanh^{-1}\beta+4(1-\ln4)\left(\frac{1}{\beta}\tanh^{-1}\beta-1\right)$ & 0\tabularnewline
			\hline 
			$K$ & $\frac{\lambda^{2}-1}{1+3\lambda^{2}}$ & $2\left(\frac{E_{e}+m_{e}}{\mathbf{p_{e}}}-\beta\right)\tanh^{-1}\beta+4(1-\ln4)\left(\frac{1}{\beta}\tanh^{-1}\beta-1\right)$ & 0\tabularnewline
			\hline 
			$L$ & $0$ & 0 & $-\frac{2\pi(\lambda^{2}-1)m_{e}}{(1+3\lambda^{2})\mathbf{p_{e}}}$\tabularnewline
			\hline 
			$N$ & $\frac{2\lambda(\lambda+1)}{1+3\lambda^{2}}\frac{m_{e}}{E_{e}}$ & $-2\beta\tanh^{-1}\beta$ & 0\tabularnewline
			\hline 
			$Q$ & $\frac{2\lambda(\lambda+1)}{1+3\lambda^{2}}$ & $2\left(\frac{E_{e}+m_{e}}{\mathbf{p_{e}}}-\beta\right)\tanh^{-1}\beta$ & 0\tabularnewline
			\hline 
			$R$ & $0$ & 0 & $\frac{4\pi\lambda(\lambda+1)m_{e}}{(1+3\lambda^{2})\mathbf{p_{e}}}$\tabularnewline
			\hline 
			$T$ & $\frac{-2\lambda(\lambda-1)}{1+3\lambda^{2}}$ & $\frac{2(1-\beta^{2})}{\beta}\tanh^{-1}\beta+4(1-\ln4)\left(\frac{1}{\beta}\tanh^{-1}\beta-1\right)$ & 0\tabularnewline
			\hline 
			$\tilde{A}$ & $\frac{2\lambda(\lambda-1)}{1+3\lambda^{2}}$ & $\frac{2(1-\beta^{2})}{\beta}\tanh^{-1}\beta$ & 0\tabularnewline
			\hline 
			$\tilde{B}$ & $\frac{-2\lambda(\lambda+1)}{1+3\lambda^{2}}$ & $4(1-\ln4)\left(\frac{1}{\beta}\tanh^{-1}\beta-1\right)$ & 0\tabularnewline
			\hline 
			$\tilde{N}$ & $\frac{-2\lambda(\lambda-1)}{1+3\lambda^{2}}\frac{m_{e}}{E_{e}}$ & $-2\beta\tanh^{-1}\beta$ & 0\tabularnewline
			\hline 
			$\tilde{Q}$ & $\frac{-2\lambda(\lambda-1)}{1+3\lambda^{2}}$ & $2\left(\frac{E_{e}+m_{e}}{\mathbf{p_{e}}}-\beta\right)\tanh^{-1}\beta$ & 0\tabularnewline
			\hline 
			$\tilde{R}$ & $0$ & 0 & $-\frac{4\pi\lambda(\lambda-1)m_{e}}{(1+3\lambda^{2})\mathbf{p_{e}}}$\tabularnewline
			\hline 
			$\tilde{T}$ & $\frac{2\lambda(\lambda+1)}{1+3\lambda^{2}}$ & $\frac{2(1-\beta^{2})}{\beta}\tanh^{-1}\beta+4(1-\ln4)\left(\frac{1}{\beta}\tanh^{-1}\beta-1\right)$ & 0\tabularnewline
			\hline 
		\end{tabular}
		\par\end{centering}
	\caption{\label{tab:RC1}Coefficients for the analytic part of the outer radiative corrections.}
\end{table}
\begin{table}
	\begin{centering}
		\begin{tabular}{|c|c|c|c|}
			\hline 
			$\mathcal{C}$ & $\mathcal{C}_{0}$ & $\delta_{\text{an},1}^{\mathcal{C}}$ & $\delta_{\text{an},2}^{\mathcal{C}}$\tabularnewline
			\hline 
			\hline 
			$X$ & $\frac{1-\lambda^{2}}{1+3\lambda^{2}}$ & 0 & 0\tabularnewline
			\hline 
			$\tilde{a}$ & $\frac{1-\lambda^{2}}{1+3\lambda^{2}}$ & $\frac{2(1-\beta^{2})}{\beta}\tanh^{-1}\beta+4(1-\ln4)\left(\frac{1}{\beta}\tanh^{-1}\beta-1\right)$ & 0\tabularnewline
			\hline 
			$\tilde{G}$ & $\frac{\lambda^{2}-1}{1+3\lambda^{2}}$ & $\frac{2(1-\beta^{2})}{\beta}\tanh^{-1}\beta$ & 0\tabularnewline
			\hline 
			$\tilde{H}$ & $\frac{\lambda^{2}-1}{1+3\lambda^{2}}\frac{m_{e}}{E_{e}}$ & $-2\beta\tanh^{-1}\beta+4(1-\ln4)\left(\frac{1}{\beta}\tanh^{-1}\beta-1\right)$ & 0\tabularnewline
			\hline 
			$\tilde{K}$ & $\frac{\lambda^{2}-1}{1+3\lambda^{2}}$ & $2\left(\frac{E_{e}+m_{e}}{\mathbf{p_{e}}}-\beta\right)\tanh^{-1}\beta+4(1-\ln4)\left(\frac{1}{\beta}\tanh^{-1}\beta-1\right)$ & 0\tabularnewline
			\hline 
			$\tilde{L}$ & 0 & 0 & $-\frac{2\pi(\lambda^{2}-1)m_{e}}{(1+3\lambda^{2})\mathbf{p_{e}}}$\tabularnewline
			\hline 
			\k{A} & $-\frac{2\lambda(\lambda-1)}{1+3\lambda^{2}}\frac{m_{e}}{E_{e}}$ & $-2\beta\tanh^{-1}\beta+4(1-\ln4)\left(\frac{1}{\beta}\tanh^{-1}\beta-1\right)$ & 0\tabularnewline
			\hline 
			\k{a} & $-\frac{2\lambda(\lambda+1)}{1+3\lambda^{2}}\frac{m_{e}}{E_{e}}$ & $-2\beta\tanh^{-1}\beta+4(1-\ln4)\left(\frac{1}{\beta}\tanh^{-1}\beta-1\right)$ & 0\tabularnewline
			\hline 
			\'{N} & $\frac{2\lambda(\lambda+1)}{1+3\lambda^{2}}$ & $\frac{2(1-\beta^{2})}{\beta}\tanh^{-1}\beta+4(1-\ln4)\left(\frac{1}{\beta}\tanh^{-1}\beta-1\right)$ & 0\tabularnewline
			\hline 
			\'{n} & $\frac{2\lambda(\lambda-1)}{1+3\lambda^{2}}$ & $\frac{2(1-\beta^{2})}{\beta}\tanh^{-1}\beta+4(1-\ln4)\left(\frac{1}{\beta}\tanh^{-1}\beta-1\right)$ & 0\tabularnewline
			\hline 
			\'{O} & 0 & 0 & $\frac{4\pi\lambda(\lambda+1)m_{e}}{(1+3\lambda^{2})\mathbf{p_{e}}}$\tabularnewline
			\hline 
			\'{o} & 0 & 0 & $\frac{4\pi\lambda(\lambda-1)m_{e}}{(1+3\lambda^{2})\mathbf{p_{e}}}$\tabularnewline
			\hline 
			\'{Z} & $-\frac{2\lambda(\lambda+1)}{1+3\lambda^{2}}$ & $2\left(\frac{E_{e}+m_{e}}{\mathbf{p_{e}}}-\beta\right)\tanh^{-1}\beta+4(1-\ln4)\left(\frac{1}{\beta}\tanh^{-1}\beta-1\right)$ & 0\tabularnewline
			\hline 
			\'{z} & $-\frac{2\lambda(\lambda-1)}{1+3\lambda^{2}}$ & $2\left(\frac{E_{e}+m_{e}}{\mathbf{p_{e}}}-\beta\right)\tanh^{-1}\beta+4(1-\ln4)\left(\frac{1}{\beta}\tanh^{-1}\beta-1\right)$ & 0\tabularnewline
			\hline 
		\end{tabular}
		\par\end{centering}
	\caption{\label{tab:RC2}(Cont.) Coefficients for the analytic part of the outer radiative corrections.}
\end{table}

\begin{table}
	\begin{centering}
		\begin{tabular}{|c|c|c|c|c|}
			\hline 
			\backslashbox{~~~~$\eta$}{$i$~~~}
			&\makebox{$a$}&\makebox{$b$}&\makebox{$c$}&\makebox{$d$}\\
			\hline 
			\hline 
			\multirow{2}{*}{1} & $E_{e}(9\lambda^{2}-4\lambda\mu_{V}+3)$ & \multirow{2}{*}{$-m_{e}(\lambda^{2}-2\lambda\mu_{V}+1)$} & $4E_{e}\lambda(\mu_{V}-3\lambda)$ & \multirow{2}{*}{$3E_{e}(\lambda^{2}-1)$}\tabularnewline
			& $+2E_{m}\lambda(\mu_{V}-\lambda)$ &  & $+2E_{m}\lambda(\lambda-\mu_{V})$ & \tabularnewline
			\hline 
			\multirow{2}{*}{2} & $E_{e}(-5\lambda^{2}+\lambda(3\mu_{V}-7)+\mu_{V})$ & \multirow{2}{*}{0} & \multirow{2}{*}{$E_{e}(\lambda+1)(5\lambda-\mu_{V})$} & \multirow{2}{*}{0}\tabularnewline
			& $+E_{m}(\lambda+1)(\lambda-\mu_{V})$ &  &  & \tabularnewline
			\hline 
			\multirow{2}{*}{3} & $E_{e}(7\lambda^{2}-\lambda(3\mu_{V}+5)+\mu_{V})$ & \multirow{2}{*}{$m_{e}(1-\lambda)(\lambda-\mu_{V})$} & $E_{e}(1-\lambda)(7\lambda-\mu_{V})$ & \multirow{2}{*}{0}\tabularnewline
			& $+2E_{m}\lambda(\mu_{V}-\lambda)$ &  & $+E_{m}(1-\lambda)(\mu_{V}-\lambda)$ & \tabularnewline
			\hline 
			\multirow{3}{*}{4} & $E_{e}(-9\lambda^{2}+4\lambda\mu_{V}-3)$ & \multirow{3}{*}{0} & $\frac{2E_{e}}{E_{e}+m_{e}}\left\{ 2E_{e}\lambda(3\lambda-\mu_{V})\right.$ & \multirow{3}{*}{$-\frac{3E_{e}^{2}(\lambda^{2}-1)}{E_{e}+m_{e}}$}\tabularnewline
			& $+2E_{m}\lambda(\lambda-\mu_{V})$ &  & $+E_{m}\lambda(\mu_{V}-\lambda)$ & \tabularnewline
			&  &  & $\left.+m_{e}(4\lambda^{2}-\lambda\mu_{V}+1)\right\} $ & \tabularnewline
			\hline 
			5 & $2m_{e}(2\lambda^{2}-\lambda\mu_{V}-1)$ & $2E_{m}\lambda(\mu_{V}-\lambda)$ & $-3m_{e}(\lambda^{2}-1)$ & $0$\tabularnewline
			\hline 
			6 & $m_{e}(\lambda+1)(5\lambda-\mu_{V})$ & $-E_{m}(\lambda+1)(\lambda-\mu_{V})$ & $m_{e}(\lambda+1)(\mu_{V}-5\lambda)$ & $0$\tabularnewline
			\hline 
			\multirow{3}{*}{7} & $E_{e}(5\lambda^{2}+\lambda(7-3\mu_{V})-\mu_{V})$ & \multirow{3}{*}{0} & \multirow{3}{*}{$E_{e}(\lambda+1)(\mu_{V}-5\lambda)$} & \multirow{3}{*}{0}\tabularnewline
			& $-E_{m}(\lambda+1)(\lambda-\mu_{V})$ &  &  & \tabularnewline
			& $-2m_{e}\lambda(\mu_{V}-1)$ &  &  & \tabularnewline
			\hline 
			\multirow{2}{*}{8} & $-E_{e}(7\lambda^{2}-\lambda(3\mu_{V}+5)+\mu_{V})$ & \multirow{2}{*}{0} & $\frac{E_{e}(\lambda-1)}{E_{e}+m_{e}}\left\{ E_{e}(7\lambda-\mu_{V})\right.$ & \multirow{2}{*}{0}\tabularnewline
			& $+2E_{m}\lambda(\lambda-\mu_{V})$ &  & $\left.+E_{m}(\mu_{V}-\lambda)+6m_{e}\lambda\right\} $ & \tabularnewline
			\hline 
			\multirow{2}{*}{9} & $E_{e}(5\lambda^{2}-\lambda(3\mu_{V}+7)+\mu_{V})$ & \multirow{2}{*}{0} & \multirow{2}{*}{$E_{e}(1-\lambda)(5\lambda-\mu_{V})$} & \multirow{2}{*}{0}\tabularnewline
			& $-E_{m}(\lambda-1)(\lambda-\mu_{V})$ &  &  & \tabularnewline
			\hline 
			\multirow{2}{*}{10} & $E_{e}(-7\lambda^{2}+\lambda(3\mu_{V}-5)+\mu_{V})$ & \multirow{2}{*}{$m_{e}(\lambda+1)(\lambda-\mu_{V})$} & $E_{e}(\lambda+1)(7\lambda-\mu_{V})$ & \multirow{2}{*}{0}\tabularnewline
			& $+2E_{m}\lambda(\lambda-\mu_{V})$ &  & $+E_{m}(\lambda+1)(\mu_{V}-\lambda)$ & \tabularnewline
			\hline 
			11 & $m_{e}(1-\lambda)(5\lambda-\mu_{V})$ & $E_{m}(\lambda-1)(\lambda-\mu_{V})$ & $m_{e}(\lambda-1)(5\lambda-\mu_{V})$ & 0\tabularnewline
			\hline 
			\multirow{3}{*}{12} & $-E_{e}(5\lambda^{2}-\lambda(3\mu_{V}+7)+\mu_{V})$ & \multirow{3}{*}{0} & \multirow{3}{*}{$E_{e}(\lambda-1)(5\lambda-\mu_{V})$} & \multirow{3}{*}{0}\tabularnewline
			& $+E_{m}(\lambda-1)(\lambda-\mu_{V})$ &  &  & \tabularnewline
			& $+2m_{e}\lambda(\mu_{V}+1)$ &  &  & \tabularnewline
			\hline 
			\multirow{2}{*}{13} & $E_{e}(7\lambda^{2}+\lambda(5-3\mu_{V})-\mu_{V})$ & \multirow{2}{*}{0} & $-\frac{E_{e}(\lambda+1)}{E_{e}+m_{e}}\left\{ E_{e}(7\lambda-\mu_{V})\right.$ & \multirow{2}{*}{0}\tabularnewline
			& $+2E_{m}\lambda(\mu_{V}-\lambda)$ &  & $\left.+E_{m}(\mu_{V}-\lambda)+6m_{e}\lambda\right\} $ & \tabularnewline
			\hline 
			14 & $3E_{e}(1-\lambda^{2})$ & $m_{e}(\lambda^{2}-1)$ & 0 & $3E_{e}(\lambda^{2}-1)$\tabularnewline
			\hline 
			15 & $3E_{e}(\lambda^{2}-1)$ & 0 & $\frac{2E_{e}m_{e}}{E_{e}+m_{e}}(1-\lambda^{2})$ & $\frac{3E_{e}^{2}(1-\lambda^{2})}{E_{e}+m_{e}}$\tabularnewline
			\hline 
		\end{tabular}
		\par\end{centering}
	\caption{\label{tab:recoil1}The recoil coefficients $r_{\eta i}$.}
\end{table}

\begin{table}
	\begin{centering}
		\begin{tabular}{|c|c|c|c|c|}
			\hline
			\backslashbox{~~~~$\eta$}{$i$~~~}
			&\makebox{$a$}&\makebox{$b$}&\makebox{$c$}&\makebox{$d$}\\ 
			\hline 
			\hline 
			16 & $2m_{e}(\lambda^{2}-1)$ & 0 & $3m_{e}(1-\lambda^{2})$ & 0\tabularnewline
			\hline 
			17 & $m_{e}(1-\lambda)(5\lambda-\mu_{V})$ & $E_{m}(\lambda-1)(\lambda-\mu_{V})$ & $6m_{e}\lambda(\lambda-1)$ & 0\tabularnewline
			\hline 
			18 & $m_{e}(1+\lambda)(\mu_{V}-5\lambda)$ & $E_{m}(\lambda+1)(\lambda-\mu_{V})$ & $6m_{e}\lambda(\lambda+1)$ & 0\tabularnewline
			\hline 
			19 & $m_{e}(1-\lambda)(\lambda+\mu_{V})$ & 0 & 0 & 0\tabularnewline
			\hline 
			20 & $-m_{e}(1+\lambda)(\lambda+\mu_{V})$ & 0 & 0 & 0\tabularnewline
			\hline 
			\multirow{2}{*}{21} & $2E_{e}(\lambda+1)(3\lambda-\mu_{V})$ & \multirow{2}{*}{0} & \multirow{2}{*}{$-6E_{e}\lambda(\lambda+1)$} & \multirow{2}{*}{0}\tabularnewline
			& $+E_{m}(\lambda+1)(\mu_{V}-\lambda)$ &  &  & \tabularnewline
			\hline 
			\multirow{2}{*}{22} & $2E_{e}(\lambda-1)(3\lambda-\mu_{V})$ & \multirow{2}{*}{0} & \multirow{2}{*}{$-6E_{e}\lambda(\lambda-1)$} & \multirow{2}{*}{0}\tabularnewline
			& $+E_{m}(\lambda-1)(\mu_{V}-\lambda)$ &  &  & \tabularnewline
			\hline 
			\multirow{2}{*}{23} & $2E_{e}(\lambda+1)(\mu_{V}-3\lambda)$ & \multirow{2}{*}{0} & \multirow{2}{*}{$6E_{e}\lambda(\lambda+1)$} & \multirow{2}{*}{0}\tabularnewline
			& $+(E_{m}-m_{e})(\lambda+1)(\lambda-\mu_{V})$ &  &  & \tabularnewline
			\hline 
			\multirow{2}{*}{24} & $2E_{e}(\lambda-1)(\mu_{V}-3\lambda)$ & \multirow{2}{*}{0} & \multirow{2}{*}{$6E_{e}\lambda(\lambda-1)$} & \multirow{2}{*}{0}\tabularnewline
			& $+(E_{m}-m_{e})(\lambda-1)(\lambda-\mu_{V})$ &  &  & \tabularnewline
			\hline 
			25 & $m_{e}(\lambda-1)(\lambda-\mu_{V})$ & $E_{m}(1-\lambda)(\lambda-\mu_{V})$ & 0 & 0\tabularnewline
			\hline 
			26 & $m_{e}(\lambda+1)(\mu_{V}-\lambda)$ & $E_{m}(1+\lambda)(\lambda-\mu_{V})$ & 0 & 0\tabularnewline
			\hline 
			27 & $-2\lambda(E_{e}-E_{m})(\lambda-\mu_{V})$ & 0 & 0 & 0\tabularnewline
			\hline 
			28 & $2\lambda(E_{e}-E_{m})(\lambda-\mu_{V})$ & 0 & 0 & 0\tabularnewline
			\hline 
			29 & $2E_{e}\lambda(\lambda+\mu_{V})$ & 0 & 0 & 0\tabularnewline
			\hline 
			30 & -$2E_{e}\lambda(\lambda+\mu_{V})$ & 0 & 0 & 0\tabularnewline
			\hline 
		\end{tabular}
		\par\end{centering}
	\caption{\label{tab:recoil2}(Cont.) The recoil coefficients $r_{\eta i}$.}
\end{table}

Now we are ready to write down the full SM prediction of the differential rate of polarized neutron decaying to polarized $p$ and $e$, including $\mathcal{O}(\alpha/\pi)$ and $\mathcal{O}(1/m_N)$ corrections. It is given in terms of the pseudo-neutrino formalism as follows:
\begin{eqnarray}
	\left(\frac{d\Gamma}{dE_ed\Omega_e d\Omega_\nu'}\right)_\text{SM}&=&\frac{\pe E_e(E_m-E_e)^2}{128\pi^5}F(E_e)\left(1+\frac{\alpha}{2\pi}\delta_\text{an}(E_e,c')\right)G_V^2(1+3\lambda^2)\nonumber\\
	&&\times \left\{1+g_\text{SM}+\frac{1}{1+3\lambda^2}\left[\frac{1}{m_N} g_\text{recoil}+\frac{\alpha}{2\pi} g_\text{brem}^\text{reg}\right]\right\}~.\label{eq:final}
\end{eqnarray}
Let us explain all the entries at the right hand side of Eq.\eqref{eq:final}:
\begin{enumerate}
	\item $F(E_e)$ is the well-known Fermi function~\cite{Fermi:1934hr}.
	\item $\delta_{\text{an}}(E_e,c')$ is the ``universal'' part of the outer radiative corrections (loop + bremsstrahlung) in the pseudo-neutrino formalism, first defined in Ref.\cite{Seng:2023ynd}:
	\begin{eqnarray}
		\delta_{\text{an}}(E_e,c')&=&-2\left(4-\ln\frac{4(E_m-E_e)^2}{m_e^2}\right)\left(\frac{1}{\beta}\tanh^{-1}\beta-1\right)+\frac{3}{2}\ln\frac{m_N^2}{m_e^2}-\frac{11}{4}\nonumber\\
		&&+2\ln\left(\frac{1-\beta c'}{1+\beta}\right)-\frac{1}{\beta}\text{Li}_2\left(\frac{2\beta}{1+\beta}\right)-\frac{1}{\beta}\text{Li}_2\left(\frac{-2\beta}{1-\beta}\right)-\frac{2}{\beta}\text{Li}_2\left(\frac{\beta(c'+1)}{1+\beta}\right)\nonumber\\
		&&+\frac{2}{\beta}\text{Li}_2\left(\frac{\beta(c'-1)}{1-\beta}\right)-\frac{2}{\beta}(\tanh^{-1}\beta)^2+2(1+\beta)\tanh^{-1}\beta~,
	\end{eqnarray}
	with $c'\equiv \cos\theta_{e\nu'}=\hat{p}_e\cdot \hat{p}_\nu'=\vec{p}_e\cdot\vec{p}_\nu'/(\pe\pnup)$, $\vec{\beta}=\pe/E_e$ and $\beta=|\vec{\beta}|$. Notice that, unlike traditional Sirlin's function, this function is angle-dependent. 
	\item $g_\text{SM}$ resembles the SM-contribution to the tree-level correlations $g_\text{JTW}+g_\text{EF}+g_{s_p}+g_{s_p s_n}$ that appear in Ref.\cite{Seng:2024fvi}:
\begin{eqnarray}
	g_\text{SM}&=&a\beta c'+\hat{s}_n\cdot\left[A\vec{\beta}+B\hat{p}_\nu'\right]+\vec{\sigma}\cdot\left[G\vec{\beta}+H\hat{p}_\nu'+K\frac{\vec{p}_e}{E_e+m_e}\beta c'+L\vec{\beta}\times\hat{p}_\nu'\right]\nonumber\\
	&&+\vec{\sigma}\cdot\left[N\hat{s}_n+Q\frac{\vec{p}_e}{E_e+m_e}\hat{s}_n\cdot\vec{\beta}+R\hat{s}_n\times\vec{\beta}\right]+T\vec{\sigma}\cdot\vec{\beta}\hat{s}_n\cdot\hat{p}_\nu'\nonumber\\
	&&+\hat{s}_p\cdot\left[\tilde{A}\vec{\beta}+\tilde{B}\hat{p}_\nu'\right]+\vec{\sigma}\cdot\left[\tilde{N}\hat{s}_p+\tilde{Q}\frac{\vec{p}_e}{E_e+m_e}\hat{s}_p\cdot\vec{\beta}+\tilde{R}\hat{s}_p\times\vec{\beta}\right]+\tilde{T}\vec{\sigma}\cdot\vec{\beta}\hat{s}_p\cdot\hat{p}_\nu'\nonumber\\
	&&+\hat{s}_p\cdot\hat{s}_n\left[X+\tilde{a}\beta c'\right]+\hat{s}_p\cdot\hat{s}_n\vec{\sigma}\cdot\left[\tilde{G}\beta+\tilde{H}\hat{p}_\nu'+\tilde{K}\frac{\vec{p}_e}{E_e+m_e}\beta c'+\tilde{L}\vec{\beta}\times\hat{p}_\nu'\right]\nonumber\\
	&&+\text{\k{A}}\vec{\sigma}\cdot\hat{s}_p\hat{s}_n\cdot\hat{p}_\nu'+\text{\k{a}}\vec{\sigma}\cdot\hat{s}_n\hat{s}_p\cdot\hat{p}_\nu'+\text{\'{N}}\hat{s}_p\cdot\hat{p}_\nu' \hat{s}_n\cdot\vec{\beta}+\text{\'{n}}\hat{s}_n\cdot\hat{p}_\nu' \hat{s}_p\cdot\vec{\beta}\nonumber\\
	&&+\text{\'{O}}\hat{s}_p\cdot\hat{p}_\nu'\vec{\sigma}\cdot\left(\vec{\beta}\times\hat{s}_n\right)+\text{\'{o}}\hat{s}_n\cdot\hat{p}_\nu'\vec{\sigma}\cdot\left(\vec{\beta}\times\hat{s}_p\right)\nonumber\\
	&&+\text{\'{Z}}\hat{s}_p\cdot\hat{p}_\nu' \hat{s}_n\cdot\vec{\beta}\frac{\sigma\cdot\vec{p}_e}{E_e+m_e}+\text{\'{z}}\hat{s}_n\cdot\hat{p}_\nu' \hat{s}_p\cdot\vec{\beta}\frac{\sigma\cdot\vec{p}_e}{E_e+m_e}~,\label{eq:gSM}
\end{eqnarray}  
{\color{black}with $\{a,A,B,G,H,K,L,N,Q,R\}$ from $g_{\text{JTW}}$, $T$ from $g_\text{EF}$, $\{\tilde{A},\tilde{B},\tilde{N},\tilde{Q},\tilde{R},\tilde{T}\}$ from $g_{s_p}$, and the remainders from $g_{s_p s_n}$. There are a few modifications:} First, $\hat{p}_\nu$ is replaced by $\hat{p}_\nu'= \vec{p}_\nu'/\pnup$ in accordance to the pseudo-neutrino formalism. Second, the correlation coefficients $\mathcal{C}=a,...,\text{\'{z}}$ in $g_\text{SM}$ are renormalized by the (non-universal) analytic part of the outer radiative corrections (loop + bremsstrahlung):
\begin{equation}
	\mathcal{C}(E_e)=\mathcal{C}_0\left(1+\frac{\alpha}{2\pi}\delta_\text{an,1}^{\mathcal{C}}(E_e)\right)+\frac{\alpha}{2\pi}\delta_\text{an,2}^{\mathcal{C}}(E_e)~.
\end{equation}
Here, $\mathcal{C}_0$ is the zeroth-order SM contribution, while $\delta_{\text{an},1}^\mathcal{C}$ and $\delta_{\text{an},2}^\mathcal{C}$ are obtained by regrouping the separate pieces that appeared in Refs.\cite{Seng:2023ynd,Seng:2024fvi}\footnote{A special case: If $\mathcal{C}_0=0$, then we may set $\delta_{\text{an},1}^\mathcal{C}=0$ for simplicity.}:
\begin{equation}
	\delta_{\text{an},1}^\mathcal{C}(E_e)=\delta_{v\text{I}}^\mathcal{C}(E_e)-\delta_{v\text{I}}^1(E_e)+\Delta_{\nu'}\delta_{\text{I}a2}(E_e)~,~\delta_{\text{an},2}^\mathcal{C}(E_e)=\frac{1}{1+3\lambda^2}\delta_{v\text{II}}^\mathcal{C}(E_e)~,
\end{equation}
where $\Delta_{\nu'}=1(0)$ if the correlation structure involves (does not involve) $\hat{p}_\nu'$. The analytic expressions of $\mathcal{C}_0$, $\delta_{\text{an},1}^\mathcal{C}$ and $\delta_{\text{an},2}^\mathcal{C}$ are provided in Table \ref{tab:RC1}-\ref{tab:RC2}. 
\item $g_\text{recoil}$ represents the $\mathcal{O}(1/m_N)$ recoil corrections, from both the squared amplitude and the phase space:
\begin{eqnarray}
    g_\text{recoil}&=&r_1+r_2\hat{s}_n\cdot \vec{\beta}+r_3\hat{
	s}_n\cdot\hat{p}_\nu'+r_4\vec{\sigma}\cdot\vec{\beta}+r_5\vec{\sigma}\cdot\hat{p}_\nu'+r_6\vec{\sigma}\cdot\hat{s}_n+r_7\hat{s}_n\cdot\vec{\beta}\frac{\vec{\sigma}\cdot\vec{p}_e}{E_e+m_e}\nonumber\\
	&&+r_8\vec{\sigma}\cdot\vec{\beta}\hat{s}_n\cdot\hat{p}_\nu'+r_9\hat{s}_p\cdot\vec{\beta}+r_{10}\hat{s}_p\cdot\hat{p}_\nu'+r_{11}\vec{\sigma}\cdot\hat{s}_p+r_{12}\hat{s}_p\cdot\vec{\beta}\frac{\vec{\sigma}\cdot\vec{p}_e}{E_e+m_e}+r_{13}\vec{\sigma}\cdot\vec{\beta}\hat{s}_p\cdot\hat{p}_\nu'\nonumber\\
	&&+r_{14}\hat{s}_n\cdot\hat{s}_p+r_{15}\hat{s}_n\cdot\hat{s}_p\vec{\sigma}\cdot\vec{\beta}+r_{16}\hat{s}_n\cdot\hat{s}_p\vec{\sigma}\cdot\hat{p}_\nu'+r_{17}\vec{\sigma}\cdot\hat{s}_p\hat{s}_n\cdot\hat{p}_\nu'+r_{18}\vec{\sigma}\cdot\hat{s}_n\hat{s}_p\cdot\hat{p}_\nu'\nonumber\\
	&&+r_{19}\vec{\sigma}\cdot\hat{s}_p\hat{s}_n\cdot\vec{\beta}+r_{20}\vec{\sigma}\cdot\hat{s}_n\hat{s}_p\cdot\vec{\beta}+r_{21}\hat{s}_p\cdot\hat{p}_\nu'\hat{s}_n\cdot\vec{\beta}+r_{22}\hat{s}_n\cdot\hat{p}_\nu'\hat{s}_p\cdot\vec{\beta}\nonumber\\
	&&+r_{23}\hat{s}_p\cdot\hat{p}_\nu' \hat{s}_n\cdot\vec{\beta}\frac{\vec{\sigma}\cdot\vec{p}_e}{E_e+m_e}+r_{24}\hat{s}_n\cdot\hat{p}_\nu' \hat{s}_p\cdot\vec{\beta}\frac{\vec{\sigma}\cdot\vec{p}_e}{E_e+m_e}+r_{25}\hat{s}_n\cdot\hat{p}_\nu'\vec{\sigma}\cdot\hat{p}_\nu'+r_{26}\hat{s}_p\cdot\hat{p}_\nu'\vec{\sigma}\cdot\hat{p}_\nu'\nonumber\\
	&&+r_{27}\hat{s}_p\cdot\hat{p}_\nu'\hat{s}_n\cdot\hat{p}_\nu'+r_{28}\hat{s}_p\cdot\hat{p}_\nu'\hat{s}_n\cdot\hat{p}_\nu'\vec{\sigma}\cdot\vec{\beta}+r_{29}\hat{s}_p\cdot\vec{\beta}\hat{s}_n\cdot\vec{\beta}+r_{30}\hat{s}_p\cdot\vec{\beta}\hat{s}_n\cdot\vec{\beta}\frac{\vec{\sigma}\cdot\vec{p}_e}{E_e+m_e}~,\nonumber\\\label{eq:grecoil}
\end{eqnarray}
where the analytic expressions of the functions
\begin{equation}
	r_\eta(E_e,c')=r_{\eta a}(E_e)+r_{\eta b}(E_e)\frac{m_e}{E_e}+r_{\eta c}(E_e)\beta c'+r_{\eta d}(E_e)\beta^2c^{\prime 2}~,~~\eta=1,...,30\label{eq:rcoeff}
\end{equation}
are derived in our supplemented \textit{Mathematica} notebook \textbf{recoil.nb} and are summarized in Tab.\ref{tab:recoil1}-\ref{tab:recoil2}. It is worthwhile to mention that, in the zeroth-order expression, an extra factor of $c'$ defines a new correlation structure (e.g. $G$ and $K$ in Eq.\eqref{eq:gSM}), but we do not do the so to $g_\text{recoil}$ otherwise Eq.\eqref{eq:grecoil} would be too long. Therefore, we classify different structures in $g_\text{recoil}$ only by their spin correlations. The same is for $g_\text{brem}^\text{reg}$ below.
\item Finally, $g_\text{brem}^\text{reg}$ encodes the contribution from the ``regular'' part of the bremsstrahlung. It takes the following form:
\begin{eqnarray}
	g_\text{brem}^\text{reg}&=&\delta^\text{reg}_1+\delta^\text{reg}_2\hat{s}_n\cdot \vec{\beta}+\delta^\text{reg}_3\hat{
		s}_n\cdot\hat{p}_\nu'+\delta^\text{reg}_4\vec{\sigma}\cdot\vec{\beta}+\delta^\text{reg}_5\vec{\sigma}\cdot\hat{p}_\nu'+\delta^\text{reg}_6\vec{\sigma}\cdot\hat{s}_n+\delta^\text{reg}_7\hat{s}_n\cdot\vec{\beta}\frac{\vec{\sigma}\cdot\vec{p}_e}{E_e+m_e}\nonumber\\
	&&+\delta^\text{reg}_8\vec{\sigma}\cdot\vec{\beta}\hat{s}_n\cdot\hat{p}_\nu'+\delta^\text{reg}_9\hat{s}_p\cdot\vec{\beta}+\delta^\text{reg}_{10}\hat{s}_p\cdot\hat{p}_\nu'+\delta^\text{reg}_{11}\vec{\sigma}\cdot\hat{s}_p+\delta^\text{reg}_{12}\hat{s}_p\cdot\vec{\beta}\frac{\vec{\sigma}\cdot\vec{p}_e}{E_e+m_e}+\delta^\text{reg}_{13}\vec{\sigma}\cdot\vec{\beta}\hat{s}_p\cdot\hat{p}_\nu'\nonumber\\
	&&+\delta^\text{reg}_{14}\hat{s}_n\cdot\hat{s}_p+\delta^\text{reg}_{15}\hat{s}_n\cdot\hat{s}_p\vec{\sigma}\cdot\vec{\beta}+\delta^\text{reg}_{16}\hat{s}_n\cdot\hat{s}_p\vec{\sigma}\cdot\hat{p}_\nu'+\delta^\text{reg}_{17}\vec{\sigma}\cdot\hat{s}_p\hat{s}_n\cdot\hat{p}_\nu'+\delta^\text{reg}_{18}\vec{\sigma}\cdot\hat{s}_n\hat{s}_p\cdot\hat{p}_\nu'\nonumber\\
	&&+\delta^\text{reg}_{19}\vec{\sigma}\cdot\hat{s}_p\hat{s}_n\cdot\vec{\beta}+\delta^\text{reg}_{20}\vec{\sigma}\cdot\hat{s}_n\hat{s}_p\cdot\vec{\beta}+\delta^\text{reg}_{21}\hat{s}_p\cdot\hat{p}_\nu'\hat{s}_n\cdot\vec{\beta}+\delta^\text{reg}_{22}\hat{s}_n\cdot\hat{p}_\nu'\hat{s}_p\cdot\vec{\beta}\nonumber\\
	&&+\delta^\text{reg}_{23}\hat{s}_p\cdot\hat{p}_\nu' \hat{s}_n\cdot\vec{\beta}\frac{\vec{\sigma}\cdot\vec{p}_e}{E_e+m_e}+\delta^\text{reg}_{24}\hat{s}_n\cdot\hat{p}_\nu' \hat{s}_p\cdot\vec{\beta}\frac{\vec{\sigma}\cdot\vec{p}_e}{E_e+m_e}+\delta^\text{reg}_{25}\hat{s}_n\cdot\hat{p}_\nu'\vec{\sigma}\cdot\hat{p}_\nu'+\delta^\text{reg}_{26}\hat{s}_p\cdot\hat{p}_\nu'\vec{\sigma}\cdot\hat{p}_\nu'\nonumber\\
	&&+\delta^\text{reg}_{27}\hat{s}_p\cdot\hat{p}_\nu'\hat{s}_n\cdot\hat{p}_\nu'+\delta^\text{reg}_{28}\hat{s}_p\cdot\hat{p}_\nu'\hat{s}_n\cdot\hat{p}_\nu'\vec{\sigma}\cdot\vec{\beta}+\delta^\text{reg}_{29}\hat{s}_p\cdot\vec{\beta}\hat{s}_n\cdot\vec{\beta}+\delta^\text{reg}_{30}\hat{s}_p\cdot\vec{\beta}\hat{s}_n\cdot\vec{\beta}\frac{\vec{\sigma}\cdot\vec{p}_e}{E_e+m_e}\nonumber\\
	&&+\delta^\text{reg}_{31}\hat{s}_p\cdot\hat{p}_\nu'\hat{s}_n\cdot\vec{\beta}\vec{\sigma}\cdot\hat{p}_\nu'+\delta^\text{reg}_{32}\hat{s}_n\cdot\hat{p}_\nu'\hat{s}_p\cdot\vec{\beta}\vec{\sigma}\cdot\hat{p}_\nu'+\delta^\text{reg}_{33}\hat{s}_p\cdot\hat{p}_\nu'\hat{s}_n\cdot\hat{p}_\nu'\vec{\sigma}\cdot\hat{p}_\nu'+\delta^\text{reg}_{34}\vec{\sigma}\cdot\hat{p}_\nu'\hat{s}_n\cdot\vec{\beta}\nonumber\\
	&&+\delta^\text{reg}_{35}\vec{\sigma}\cdot\hat{p}_\nu'\hat{s}_p\cdot\vec{\beta}+\delta^\text{reg}_{36}\hat{s}_n\cdot\vec{\beta}\hat{s}_p\cdot\vec{\beta}\vec{\sigma}\cdot\hat{p}_\nu'~,\label{eq:gbremreg}
\end{eqnarray}
where the functions $\delta^\text{reg}_\eta=\delta^\text{reg}_\eta(E_e,c')$ ($\eta=1,...,36$) result from the integration of Eq.\eqref{eq:M4reg} over $p_\nu$, $k$ and $\pnup$. They are too complicated to be displayed analytically, but are evaluated numerically in the supplemented \textit{Mathematica} notebook \textbf{bremreg.nb}.
\end{enumerate}

\section{\label{sec:check}Comparing to existing literature}

Since we claim to have derived the SM prediction to the most general neutron differential decay rate, an important step is to compare our result to all special cases available in existing literature. We do this for both the recoil and radiative corrections.  

\subsection{Recoil corrections} 

Since the $\mathcal{O}(1/m_N)$ recoil corrections involve only the 3-body decay process, there is no difference between $\hat{p}_\nu$ and $\hat{p}_\nu'$ and we can directly compare our result with existing literature for the differential decay rare, either with integrated or unintegrated $\Omega_\nu$.

\subsubsection{Polarized $n$ and unpolarized $p$, $e$}

Cases with polarized neutron and unpolarized proton and electron are most frequently studied. We compare our $r_1$--$r_3$ with the expression of $d\Gamma/(dE_ed\Omega_ed\Omega_\nu)$ in, e.g. Refs.\cite{Ando:2004rk,Gudkov:2005bu,Bhattacharya:2011qm,Cirigliano:2022hob} (notice: $\lambda_\text{there}=-\lambda_\text{here}$), which we find perfect agreement\footnote{A typo is present in the expression of $c_1^{(A)}$ in the supplementary material of Ref.\cite{Cirigliano:2022hob}, while Ref.\cite{Bhattacharya:2011qm} has the correct expression.} .    

\subsubsection{Polarized $e$ and unpolarized $n$, $p$}

We compare our $r_4$, $r_5$ with the expression of $d\Gamma/(dE_ed\Omega_ed\Omega_\nu)$ in Ref.\cite{Ivanov:2018yir}, which we find perfect agreement.    

\subsubsection{Polarized $n$, $e$ and unpolarized $p$}

Recoil corrections to correlations that involve the simultaneous polarization of neutron and electron are studied in Refs.\cite{Ivanov:2017mnz,Ivanov:2018vmz}. These references focused on the differential decay rate with integrated neutrino solid angle, $d\Gamma/(dE_ed\Omega_e)$. To compare with our result, we need to perform the $\Omega_\nu'$ integration:
\begin{equation}
	\frac{d\Gamma}{dE_ed\Omega_e}=\int d\Omega_\nu'\frac{d\Gamma}{dE_ed\Omega_ed\Omega_\nu'}=\int d\Omega_\nu\frac{d\Gamma}{dE_ed\Omega_ed\Omega_\nu}~,\label{eq:dGammadE}
\end{equation}
where the following identities are useful in dealing with the azimuthal angle in $\Omega_\nu'$:
\begin{eqnarray}
	\int d\Omega_\nu'\vec{s}\cdot \hat{p}_\nu'&=&\int d\Omega_\nu' \frac{c'}{\beta}\vec{s}\cdot\vec{\beta}\nonumber\\
	\int d\Omega_\nu'\vec{s}_1\cdot \hat{p}_\nu'\vec{s}_2\cdot\hat{p}_\nu'&=&\int d\Omega_\nu'\left[\frac{3c^{\prime 2}-1}{2\beta^2}\vec{s}_1\cdot\vec{\beta}\vec{s}_2\cdot\vec{\beta}+\frac{1-c^{\prime 2}}{2}\vec{s}_1\cdot\vec{s}_2\right]\nonumber\\
	\int d\Omega_\nu'\vec{s}_1\cdot\hat{p}_\nu' \vec{s}_2\cdot\hat{p}_\nu' \vec{s}_3\cdot\hat{p}_\nu'&=&\int d\Omega_\nu'\:c'\left[\frac{5c^{\prime 2}-3}{2\beta^3}\vec{s}_1\cdot\vec{\beta}\vec{s}_2\cdot\vec{\beta}\vec{s}_3\cdot\vec{\beta}\right.\nonumber\\
	&&\left.+\frac{1-c^{\prime 2}}{2\beta}\left(\vec{s}_1\cdot\vec{\beta}\vec{s}_2\cdot \vec{s}_3+\vec{s}_2\cdot\vec{\beta}\vec{s}_1\cdot \vec{s}_3+\vec{s}_3\cdot\vec{\beta}\vec{s}_1\cdot \vec{s}_2\right)\right]~.\label{eq:azimuthal}
\end{eqnarray}
Using this we obtain, for the $\hat{s}_n\cdot\vec{\sigma}$ structure, the following identity:
\begin{eqnarray}
	\frac{1}{2}\int_{-1}^{1}dc'\left(r_6+\frac{1-c^{\prime 2}}{2}r_{25}\right)&=&\frac{m_e}{E_e}\left[\left(\frac{16}{3}\lambda^2-\left(\frac{4}{3}\kappa_V-\frac{10}{3}\right)\lambda-\frac{2}{3}(\kappa_V+1)\right)E_e\right.\nonumber\\
	&&\left.-\left(\frac{4}{3}\lambda^2-\left(\frac{4}{3}\kappa_V+\frac{2}{3}\right)\lambda-\frac{2}{3}(\kappa_V+1)\right)E_m\right]~.
\end{eqnarray}
Meanwhile, for the $\hat{s}_n\cdot\vec{\beta}\vec{\sigma}\cdot\vec{\beta}$ structure we obtain the following identity:
\begin{eqnarray}
	\frac{1}{2}\int_{-1}^{1}dc'\left(r_7+\frac{E_e+m_e}{E_e}\left[\frac{c'}{\beta}r_8+\frac{3c^{\prime 2}-1}{2\beta^2}r_{25}\right]\right)&=&\left(\frac{22}{3}\lambda^2-\left(\frac{10}{3}\kappa_V-\frac{4}{3}\right)\lambda-\frac{2}{3}(\kappa_V+1)\right)E_e\nonumber\\
	&&-\left(\frac{4}{3}\lambda^2-\left(\frac{4}{3}\kappa_V+\frac{2}{3}\right)\lambda-\frac{2}{3}(\kappa_V+1)\right)E_m\nonumber\\
	&&+\left(2\lambda^2-2(\kappa_V+1)\lambda\right)m_e~.
\end{eqnarray}
The right hand side of these two equations has a slight difference in the $\mathcal{O}(\lambda)$ terms from the corresponding expressions in the aforementioned references (e.g. Eq.(7) in Ref.\cite{Ivanov:2017mnz}) which, we believe, indicate typos in the latter. Our full derivation of the recoil corrections is provided in \textbf{recoil.nb} so interested readers can easily check its correctness. 

\subsection{Radiative corrections\label{sec:checkRC}}

The main obstacle in comparing our result of the radiative corrections to existing literature based on the neutrino formalism is that  $\Omega_\nu'\neq\Omega_\nu$, so
\begin{equation}
	\frac{d\Gamma}{dE_ed\Omega_ed\Omega_\nu'}\neq \frac{d\Gamma}{dE_ed\Omega_ed\Omega_\nu}~.
\end{equation}
Fortunately, the two formalisms have to reconcile once the respective neutrino or pseudo-neutrino solid angle is integrated out, as indicated in Eq.\eqref{eq:dGammadE}; that allows us to at least compare our results with literature for the $\Omega_\nu'$-integrated expression, again making use of Eq.\eqref{eq:azimuthal} for the azimuthal angles. We adopt the notations by Ivanov \textit{et al.} (cited below) for the various functions for the outer corrections in the neutrino formalism. 

\subsubsection{Spin-independent structure}

For this structure, we checked numerically that our result satisfies the following identity:
\begin{equation}
	g_n(E_e)=\frac{1}{4}\int_{-1}^{1}dc'\left[\delta_{\text{an}}(E_e,c')+\frac{1-\lambda^2}{1+3\lambda^2}\beta c'\left(\delta_{\text{an}}(E_e,c')+\delta_{\text{an},1}^a(E_e)\right)+\frac{\delta_1^\text{reg}(E_e,c')}{1+3\lambda^2}\right]~,
\end{equation}
where $g_n(E_e)$ is one-half of the well-known Sirlin's function $g(E_e)$~\cite{Sirlin:1967zza}.

\subsubsection{The $\hat{s}_n\cdot\vec{\beta}$ structure}

For this structure, we checked numerically that our result satisfies the following identity:
\begin{eqnarray}
-\frac{2\lambda(\lambda+1)}{1+3\lambda^2}\left(g_n(E_e)+f_n(E_e)\right)&=&\frac{1}{4}\int_{-1}^{1}dc'\left[-\frac{2\lambda(\lambda+1)}{1+3\lambda^2}\left(\delta_{\text{an}}(E_e,c')+\delta_{\text{an},1}^A(E_e)\right)\right.\nonumber\\
&&+\frac{2\lambda(\lambda-1)}{1+3\lambda^2}\frac{c'}{\beta}\left(\delta_{\text{an}}(E_e,c')+\delta_{\text{an},1}^B(E_e)\right)+\frac{\delta_2^\text{reg}(E_e,c')}{1+3\lambda^2}\nonumber\\
&&\left.+\frac{c'}{\beta}\frac{\delta_3^\text{reg}(E_e,c')}{1+3\lambda^2}\right]~,
\end{eqnarray}
where $f_n(E_e)$ can be found, e.g. in Eq.(A9) of Ref.\cite{Ivanov:2017mnz}. These two identities had in fact already been verified in our previous paper, Ref.\cite{Seng:2023ynd}.

\subsubsection{The $\vec{\sigma}\cdot\vec{\beta}$ structure}

For this structure, we checked numerically that our result satisfies the following identity:
\begin{eqnarray}
	-\left(g_n(E_e)+f_n(E_e)\right)&=&\frac{1}{4}\int_{-1}^{1}dc'\left[-\left(\delta_{\text{an}}(E_e,c')+\delta_{\text{an},1}^G(E_e)\right)+\frac{\lambda^2-1}{1+3\lambda^2}\frac{m_e}{E_e}\frac{c'}{\beta}\left(\delta_{\text{an}}(E_e,c')+\delta_{\text{an},1}^H(E_e)\right)\right.\nonumber\\
	&&\left.+\frac{\lambda^2-1}{1+3\lambda^2}\frac{E_e}{E_e+m_e}\beta c'\left(\delta_{\text{an}}(E_e,c')+\delta_{\text{an},1}^K(E_e)\right)+\frac{\delta_4^\text{reg}(E_e,c')}{1+3\lambda^2}+\frac{c'}{\beta}\frac{\delta_5^\text{reg}(E_e,c')}{1+3\lambda^2}\right]~.\nonumber\\
\end{eqnarray}

\subsubsection{The $\hat{s}_n\cdot\vec{\sigma}$ structure}

For this structure, we checked numerically that our result satisfies the following identity:
\begin{eqnarray}
	\frac{2\lambda(\lambda+1)}{1+3\lambda^2}\frac{m_e}{E_e}\left(g_n(E_e)+h_n^{(1)}(E_e)\right)&=&\frac{1}{4}\int_{-1}^{1}dc'\left[\frac{2\lambda(\lambda+1)}{1+3\lambda^2}\frac{m_e}{E_e}\left(\delta_{\text{an}}(E_e,c')+\delta_{\text{an},1}^N(E_e)\right)\right.\nonumber\\
	&&\left.+\frac{\delta_6^\text{reg}(E_e,c')}{1+3\lambda^2}+\frac{1-c^{\prime 2}}{2}\frac{\delta_{25}^\text{reg}(E_e,c')}{1+3\lambda^2}\right]~,\nonumber\\
\end{eqnarray}
where the correct analytic expression of $h_n^{(1)}(E_e)$ (and $h_n^{(2)}(E_e)$ below) is given in the Erratum of Ref.\cite{Ivanov:2017mnz}.

\subsubsection{The $\hat{s}_n\cdot\vec{\beta}\vec{\sigma}\cdot\vec{\beta}$ structure}

For this structure, we checked numerically that our result satisfies the following identity:
\begin{eqnarray}
	\frac{2\lambda(\lambda+1)}{1+3\lambda^2}\frac{E_e}{E_e+m_e}\left(g_n(E_e)+h_n^{(2)}(E_e)\right)&=&\frac{1}{4}\int_{-1}^{1}dc'\left[\frac{2\lambda(\lambda+1)}{1+3\lambda^2}\frac{E_e}{E_e+m_e}\left(\delta_{\text{an}}(E_e,c')+\delta_{\text{an},1}^Q(E_e)\right)\right.\nonumber\\
	&&-\frac{2\lambda(\lambda-1)}{1+3\lambda^2}\frac{c'}{\beta}\left(\delta_{\text{an}}(E_e,c')+\delta_{\text{an},1}^T(E_e)\right)\nonumber\\
	&&+\frac{E_e}{Ee+m_e}\frac{\delta_7^\text{reg}(E_e,c')}{1+3\lambda^2}+\frac{c'}{\beta}\frac{\delta_{8}^\text{reg}(E_e,c')}{1+3\lambda^2}\nonumber\\
	&&\left.+\frac{3c^{\prime 2}-1}{2\beta^2}\frac{\delta_{25}^\text{reg}(E_e,c')}{1+3\lambda^2}+\frac{c'}{\beta}\frac{\delta_{34}^\text{reg}(E_e,c')}{1+3\lambda^2}\right]~.
\end{eqnarray}
Thus, we have checked that our results of outer radiative corrections are consistent to all known results in literature.

\section{\label{sec:summary}Summary}

This work provides a solid theory foundation for the proposal in Ref.\cite{Seng:2024fvi} to measure experimentally the proton polarization in the free neutron decay. We computed the SM-induced, $\mathcal{O}(1/m_N)$ recoil corrections and $\mathcal{O}(\alpha/\pi)$ radiative corrections to the most general neutron differential decay rate, with $n$, $p$ and $e$ all polarized; the former is fully analytic, while the latter is analytic apart from a regular $\pnup$-integration which is performed numerically. Our choice of independent variables ($E_e,\Omega_e,\Omega_\nu'$) are all measurable quantities in experiments with no ambiguity caused by emissions of extra photons. We compare our results to special cases in existing literature and identify possible typos in the latter. 
Together with the EFT analysis in Ref.\cite{Seng:2024fvi}, it opens a new window for the precision test of the SM and the search for new physics.

\begin{acknowledgments}
		
The work of C.-Y.S. is supported in
part by the U.S. Department of Energy (DOE), Office of Science, Office of Nuclear Physics, under the FRIB Theory Alliance award DE-SC0013617, and by the DOE grant DE-FG02-97ER41014. We acknowledge support from the DOE Topical Collaboration ``Nuclear Theory for New Physics'', award No. DE-SC0023663. 
		
\end{acknowledgments}
	
\begin{appendix}
	
\section{\label{sec:scalar}Scalar integrals in the bremsstrahlung contribution}

An important step to evaluate the ``regular'' bremsstrahlung contribution to the differential decay rate is to compute analytically the phase space integration of $|\mathcal{M}_4|_\text{QM,reg}^2$ (see Eq.\eqref{eq:M4reg}) with respect to the momenta $p_\nu$ and $k$. For terms in the squared amplitude independent of various spin vectors, this corresponds to evaluating scalar integrals of the following form ($p_1=p$, $p_2=p_e$):
\begin{equation}
	I_{i,j}(p_1,p_2)\equiv \int\frac{d^3k}{(2\pi)^32E_k}\frac{d	^3p_\nu}{(2\pi)^32E_\nu}(2\pi)^4\delta^{(4)}(p_\nu'-p_\nu-k)\frac{1}{(p_1\cdot k)^i(p_2\cdot k)^j}~,
\end{equation}
where $i$, $j$ are integers. In the next Appendix we show that integrations of terms with spin vectors can also be reduced to such scalar integrals. Ref.\cite{Ginsberg:1969jh} gave the first analytic expressions of these integrals for values of $\{i,j\}$ relevant to the study of semileptonic kaon decays (beware of the the difference in normalization with this paper), which are also quoted in Ref.\cite{Seng:2023ynd}. For beta decays involving two or more polarizations, we need one more integral, namely $I_{-3,2}$, which we present here for the first time. Here we summarize the integrals that we need (the other half can be obtained by $I_{i,j}(p_1,p_2)=I_{j,i}(p_2,p_1)$):
\begin{eqnarray}
	I_{0,0}(p_1,p_2)&=&\frac{1}{8\pi}\\
	I_{-1,0}(p_1,p_2)&=&\frac{\alpha_1}{16\pi}\\
	I_{1,0}(p_1,p_2)&=&\frac{1}{8\pi\beta_1}\ln\frac{\alpha_1+\beta_1}{\alpha_1-\beta_1}\\
	I_{2,0}(p_1,p_2)&=&\frac{1}{2\pi m_1^2 p_\nu^{\prime 2}}\\
	I_{1,1}(p_1,p_2)&=&\frac{1}{4\pi\gamma_{12}p_\nu^{\prime 2}}\ln\frac{p_1\cdot p_2+\gamma_{12}}{p_1\cdot p_2-\gamma_{12}}\\
	I_{1,-1}(p_1,p_2)&=&\frac{1}{8\pi}\left(\frac{(p_1p_2:p_\nu')}{\beta_1^2}+\frac{p_\nu^{\prime 2}(p_2p_\nu':p_1)}{2\beta_1^3}\ln\frac{\alpha_1+\beta_1}{\alpha_1-\beta_1}\right)\\
	I_{2,-1}(p_1,p_2)&=&\frac{1}{8\pi}\left(\frac{2(p_2p_\nu':p_1)}{m_1^2\beta_1^2}+\frac{(p_1p_2:p_\nu')}{\beta_1^3}\ln\frac{\alpha_1+\beta_1}{\alpha_1-\beta_1}\right)\\
	I_{-2,1}(p_1,p_2)&=&\frac{1}{8\pi}\left[\frac{(p_\nu^{\prime 2})^2(p_1p_\nu':p_2)^2}{4\beta_2^5}\ln\frac{\alpha_2+\beta_2}{\alpha_2-\beta_2}+\frac{p_\nu^{\prime 2}(p_2p_1:p_\nu')(p_1p_\nu':p_2)}{\beta_2^4}+\frac{\alpha_2(p_2p_1:p_\nu')^2}{2\beta_2^4}\right.\nonumber\\
	&&\left.+\left(\frac{\beta_2^2\beta_1^2-(p_2p_1:p_\nu')^2}{4\beta_2^4}\right)\left(\alpha_2-\frac{m_2^2p_\nu^{\prime 2}}{2\beta_2}\ln\frac{\alpha_2+\beta_2}{\alpha_2-\beta_2}\right)\right]\\
	I_{-2,2}(p_1,p_2)&=&\frac{1}{8\pi}\left[\frac{p_\nu^{\prime 2}(p_1p_\nu':p_2)^2}{m_2^2\beta_2^4}+\frac{p_\nu^{\prime 2}(p_1p_\nu':p_2)(p_2p_1:p_\nu')}{\beta_2^5}\ln\frac{\alpha_2+\beta_2}{\alpha_2-\beta_2}+\frac{(p_2p_1:p_\nu')^2}{\beta_2^4}\right.\nonumber\\
	&&\left.-\left(\frac{\beta_2^2\beta_1^2-(p_2p_1:p_\nu')^2}{2\beta_2^4}\right)\left(2-\frac{\alpha_2}{\beta_2}\ln\frac{\alpha_2+\beta_2}{\alpha_2-\beta_2}\right)\right]\\
	I_{-3,2}(p_1,p_2)&=&\frac{1}{64\pi\beta_2^7}\left\{3\left[2\alpha_1^2\beta_2^4\delta+2\alpha_1\alpha_2\beta_2^2(\beta_1^2\beta_2^2-3\delta^2)+\alpha_2^2(5\delta^3-3\beta_1^2\beta_2^2\delta)\right.\right.\nonumber\\
	&&\left.+\beta_1^2\beta_2^4\delta-\beta_2^2\delta^3\right]\ln\frac{\alpha_2+\beta_2}{\alpha_2-\beta_2}+\frac{2}{\alpha_2^2-\beta_2^2}\left[2\alpha_1^3\beta_2^7-6\alpha_1^2\alpha_2\beta_2^5\delta\right.\nonumber\\
	&&+6\alpha_1(\alpha_2^2(3\beta_2^3\delta^2-\beta_1^2\beta_2^5)+\beta_1^2\beta_2^7-2\beta_2^5\delta^2)\nonumber\\
	&&\left.\left.+\alpha_2\beta_2\delta(3\alpha_2^2(3\beta_1^2\beta_2^2-5\delta^2)-9\beta_1^2\beta_2^4+13\beta_2^2\delta^2)\right]\right\}
\end{eqnarray}
where we have defined:
\begin{eqnarray}
	\alpha_i&\equiv p_i\cdot p_\nu'\nonumber\\
	\beta_i&\equiv&\sqrt{\alpha_i^2-m_i^2p_\nu^{\prime 2}}\nonumber\\
	\gamma_{12}&\equiv&\sqrt{(p_1\cdot p_2)^2-m_1^2m_2^2}\nonumber\\
	(ab:c)&\equiv&(a\cdot c)(b\cdot c)-c^2(a\cdot b)
\end{eqnarray}
and $\delta\equiv (p_1p_2:p_\nu')$ in $I_{-3,2}$ for simplicity.
		
\section{\label{sec:tensor}Tensor integrals in the regular bremsstrahlung contribution}

If there are polarized massive particles, then we also need to deal with integrals that contain extra factors of $k^\mu$ (up to 3 for neutron beta decay) in the numerator, which will be later contracted to the spin vectors. In this Appendix we outline the systematic approach to evaluate these tensor integrals.  		

\subsection{With $k^\mu$}

If one out of the three massive particles is polarized, then we may encounter vector integrals of the following form:
\begin{equation}
	I_{i,j}^\mu\equiv \int\frac{d^3k}{(2\pi)^32E_k}\frac{d	^3p_\nu}{(2\pi)^32E_\nu}(2\pi)^4\delta^{(4)}(p_\nu'-p_\nu-k)\frac{k^\mu}{(p\cdot k)^i(p_e\cdot k)^j}~.
\end{equation}
They were first studied in Ref.\cite{Seng:2023ynd} , which we will briefly recap here to introduce further discussions. One starts by the following general decomposition of the vector integral:
\begin{equation}
     I_{i,j}^\mu\equiv a_1(i,j)p^\mu+a_2(i,j)p_e^\mu+a_3(i,j)p_\nu^{\prime\mu}~.
\end{equation}
The functions $a_1$, $a_2$, $a_3$ can be solved by contracting both sides by $p_\mu$, $p_{e\mu}$ and $(p_\nu')_\mu$ respectively, which gives rise to the following matrix equation:
\begin{equation}
	\left(\begin{array}{c}
		a_{1}(i,j)\\
		a_{2}(i,j)\\
		a_{3}(i,j)
	\end{array}\right)=M^{-1}\left(\begin{array}{c}
		I_{i-1,j}\\
		I_{i,j-1}\\
		\frac{p_{\nu}^{\prime2}}{2}I_{i,j}
	\end{array}\right)~,
\end{equation} 
where 
\begin{equation}
	M=\left(\begin{array}{ccc}
		m_{N}^{2} & p\cdot p_{e} & p\cdot p_{\nu}'\\
		p\cdot p_{e} & m_{e}^{2} & p_{e}\cdot p_{\nu}'\\
		p\cdot p_{\nu}' & p_{e}\cdot p_{\nu}' & p_{\nu}^{\prime2}
	\end{array}\right)\label{eq:Mmatrix}
\end{equation}
is a $3\times 3$ matrix of which inverse can be easily computed. Here we have used the identity $p_\nu'\cdot k=p_\nu^{\prime 2}/2$. 

\subsection{With $k^\mu k^\nu$}

When there are two polarized particles, we need also the following tensor integral 
\begin{equation}
	I_{i,j}^{\mu\nu}\equiv \int\frac{d^3k}{(2\pi)^32E_k}\frac{d	^3p_\nu}{(2\pi)^32E_\nu}(2\pi)^4\delta^{(4)}(p_\nu'-p_\nu-k)\frac{k^\mu k^\nu}{(p\cdot k)^i(p_e\cdot k)^j}~.
\end{equation}
Again, we adopt the most general decomposition:
\begin{eqnarray}
	I_{i,j}^{\mu\nu}&=&b_1(i,j)g^{\mu\nu}+b_2(i,j)p^\mu p^\nu+b_3(i,j)p_e^\mu p_e^\nu+b_4(i,j)p_\nu^{\prime\mu}p_\nu^{\prime\nu}+b_5(i,j)(p^\mu p_e^\nu+p^\nu p_e^\mu)\nonumber\\
	&&+b_6(i,j)(p^\mu p_\nu^{\prime\nu}+p^\nu p_\nu^{\prime\mu})+b_7(i,j)(p_e^\mu p_\nu^{\prime\nu}+p_e^\nu p_\nu^{\prime\mu})~.\label{eq:Imuij}
\end{eqnarray}
In principle, one may solve the functions $b_1$--$b_7$ using exactly the same method, but that will involve the inversion of a $7\times 7$ matrix which is analytically very challenging, and the numerical inversion of a badly conditioned matrix may also lead to significant errors. 

Here we introduce a trick to obtain the analytic expressions of all these coefficients without inverting a large matrix. We consider the following singly-contracted integral:
\begin{equation}
	\bar{I}^\mu_{i,j}\equiv s_\nu I^{\mu\nu}_{i,j}~,
\end{equation} 
where $s_\mu$ is an arbitrary vector. The most general decomposition of this integral reads:
\begin{equation}
	\bar{I}^{\mu}_{i,j}=\bar{b}_1(i,j)p^\mu+\bar{b}_2(i,j)p_e^\mu+\bar{b}_3(i,j)p_\nu^{\prime\mu}+b_1(i,j)s^\mu~.\label{eq:Ibarmuij}
\end{equation}
In particular, the coefficient of the $s^\mu$ term is the same $b_1(i,j)$ that appears in Eq.\eqref{eq:Imuij}, which is obvious by contracting $p_\nu$ to both sides of that equation; so we may start by evaluating $b_1(i,j)$. For that purpose, it is useful to define a set of unit vectors from the available vectors in the integral, namely $p^\mu$, $p_e^\mu$ and $p_\nu^{\prime\mu}$. First, the temporal unit vector reads:
\begin{equation}
	\hat{t}^\mu\equiv (1,\vec{0})=\frac{1}{m_N}p^\mu~.\label{eq:hattmu}
\end{equation}
Next, we arrange the coordinates such that $\vec{p}_e$ aligns with the $x$-axis. With that we define:
\begin{equation}
	\hat{x}^\mu\equiv (0,\hat{x})=\frac{1}{\pe}(p_e^\mu-E_e \hat{t}^\mu)=\frac{1}{\pe}\left(p_e^\mu-\frac{E_e}{m_N}p^\mu\right)~.\label{eq:hatxmu}
\end{equation}
Finally, we align $\vec{p}_\nu'$ to contain only $x$- and $y$-components. With that we define:
\begin{equation}
	\hat{y}^\mu\equiv (0,\hat{y})=\frac{1}{\pnup s'}(p_\nu^{\prime\mu}-E_\nu'\hat{t}^\mu-\pnup c'\hat{x}^\mu)=\frac{1}{\pnup s'}\left(p_\nu^{\prime\mu}-\frac{E_\nu^\prime \pe-\pnup E_e c'}{\pe m_N}p^\mu-\frac{\pnup c'}{\pe}p_e^\mu\right)~,\label{eq:hatymu}
\end{equation}		
where $s'\equiv \sin\theta_{e\nu'}$. 

It is easy to see that 
\begin{equation}
I_{i,j}^{zz}=b_1(i,j)g^{33}=-b_1(i,j)\label{eq:Iijb1}
\end{equation}
because none of the vectors $p$, $p_e$, $p_\nu'$ has a $z$-component. At the same time, the on-shell momentum of the bremsstrahlung photon satisfies:
\begin{eqnarray}
	-(k^z)^2&=&-(k^0)^2+(k^x)^2+(k^y)^2\nonumber\\
	&=&-(k\cdot\hat{t})^2+(k\cdot\hat{x})^2+(k\cdot\hat{y})^2\nonumber\\
	&\equiv&b_{1a}+b_{1b}p\cdot k+b_{1c}p_e\cdot k+b_{1d}(p\cdot k)^2+b_{1e}(p_e\cdot k)^2+b_{1f}(p\cdot k)(p_e\cdot k)~,\label{eq:kz2}
\end{eqnarray} 
where the coefficients $\{b_{1a},...,b_{1f}\}$ are deducible from Eqs.\eqref{eq:hattmu}--\eqref{eq:hatymu}. Plugging this back into Eq.\eqref{eq:Iijb1} gives:
\begin{equation}
	b_1(i,j)=b_{1a}I_{i,j}+b_{1b}I_{i-1,j}+b_{1c}I_{i,j-1}+b_{1d}I_{i-2,j}+b_{1e}I_{i,j-2}+b_{1f}I_{i-1,j-1}~.\label{eq:b1decompose}
\end{equation}

With $b_1$ determined, we can now deduce the the three remaining coefficients in Eq.\eqref{eq:Ibarmuij}. Moving the $b_1 s^\mu$ term to the left and contracting both sides by $p_\mu$, $p_{e\mu}$ and $(p_\nu')_\mu$, we obtain the following matrix equation:
\begin{equation}
	\left(\begin{array}{c}
		\bar{b}_{1}(i,j)\\
		\bar{b}_{2}(i,j)\\
		\bar{b}_{3}(i,j)
	\end{array}\right)=M^{-1}\left(\begin{array}{c}
		p_{\mu}\bar{I}_{i,j}^{\mu}-b_{1}(i,j)s\cdot p\\
		p_{e\mu}\bar{I}_{i,j}^{\mu}-b_{1}(i,j)s\cdot p_{e}\\
		(p_{\nu}')_{\mu}\bar{I}_{i,j}^{\mu}-b_{1}(i,j)s\cdot p_{\nu}'
	\end{array}\right)=M^{-1}\left(\begin{array}{c}
		s_{\mu}I_{i-1,j}^{\mu}-b_{1}(i,j)s\cdot p\\
		s_{\mu}I_{i,j-1}^{\mu}-b_{1}(i,j)s\cdot p_{e}\\
		\frac{p_{\nu}^{\prime2}}{2}s_{\mu}I_{i,j}^{\mu}-b_{1}(i,j)s\cdot p_{\nu}'
	\end{array}\right)~.
\end{equation}
Notice that there is no new matrix inversion required in this step apart from the known $M^{-1}$ from the previous subsection. 
We then plug the solution back into Eq.\eqref{eq:Ibarmuij} and remove $s_\mu$ from both sides because it is an arbitrary vector. That gives:
\begin{eqnarray}
	I_{i,j}^{\mu\nu}&=&b_1(i,j)g^{\mu\nu}+\left\{M_{11}^{-1}(I_{i-1,j}^\nu-b_1(i,j)p^\nu)+M_{12}^{-1}(I_{i,j-1}^\nu-b_1(i,j)p_e^\nu)\right.\nonumber\\
	&&\left.+M_{13}^{-1}((p_\nu^{\prime 2}/2)I_{i,j}^\nu-b_1(i,j)(p_\nu')^\nu)\right\}p^\mu+\left\{M_{21}^{-1}(I_{i-1,j}^\nu-b_1(i,j)p^\nu)\right.\nonumber\\
	&&\left.+M_{22}^{-1}(I_{i,j-1}^\nu-b_1(i,j)p_e^\nu)+M_{23}^{-1}((p_\nu^{\prime 2}/2)I_{i,j}^\nu-b_1(i,j)(p_\nu')^\nu)\right\}p_e^\mu\nonumber\\
	&&+\left\{M_{31}^{-1}(I_{i-1,j}^\nu-b_1(i,j)p^\nu)+M_{32}^{-1}(I_{i,j-1}^\nu-b_1(i,j)p_e^\nu)\right.\nonumber\\
	&&\left.+M_{33}^{-1}((p_\nu^{\prime 2}/2)I_{i,j}^\nu-b_1(i,j)(p_\nu')^\nu)\right\}(p_\nu')^\mu~.
\end{eqnarray}
Comparing this to Eq.\eqref{eq:Imuij} yields all the quantities $b_\eta(i,j)$, which can be decomposed similar to Eq.\eqref{eq:b1decompose} as:	
\begin{equation}
	b_\eta(i,j)=b_{\eta a}I_{i,j}+b_{\eta b}I_{i-1,j}+b_{\eta c}I_{i,j-1}+b_{\eta d}I_{i-2,j}+b_{\eta e}I_{i,j-2}+b_{\eta f}I_{i-1,j-1}~,~\eta=1,...,7~.\label{eq:bexpand}
\end{equation}		
The analytic expressions of the coefficients $\{b_{\eta a},...,b_{\eta f}\}$ ($\eta=1,...,7$) can be found in the supplemented \textit{Mathematica} notebook \textbf{bremreg.nb}.
		
\subsection{With $k^\mu k^\nu k^\alpha$}		

When $n$, $p$ and $e$ are all polarized, we may encounter tensor integrals of the form:
\begin{equation}
	I_{i,j}^{\mu\nu\alpha}\equiv \int\frac{d^3k}{(2\pi)^32E_k}\frac{d	^3p_\nu}{(2\pi)^32E_\nu}(2\pi)^4\delta^{(4)}(p_\nu'-p_\nu-k)\frac{k^\mu k^\nu k^\alpha}{(p\cdot k)^i(p_e\cdot k)^j}~,
\end{equation}
which adopts the following general decomposition:
\begin{eqnarray}
	I_{i,j}^{\mu\nu\alpha}&=&c_1(i,j)(g^{\mu\nu}p^\alpha+g^{\mu\alpha}p^\nu+g^{\nu\alpha}p^\mu)+c_2(i,j)(g^{\mu\nu}p_e^\alpha+g^{\mu\alpha}p_e^\nu+g^{\nu\alpha}p_e^\mu)\nonumber\\
	&&+c_3(i,j)(g^{\mu\nu}(p_\nu')^\alpha+g^{\mu\alpha}(p_\nu')^\nu+g^{\nu\alpha}(p_\nu')^\mu)+c_4(i,j)p^\mu p^\nu p^\alpha+c_5(i,j)p_e^\mu p_e^\nu p_e^\alpha\nonumber\\
	&&+c_6(i,j)(p_\nu')^\mu(p_\nu')^\nu(p_\nu')^\alpha+c_7(i,j)(p^\mu p^\nu p_e^\alpha+p^\mu p^\alpha p_e^\nu+p^\nu p^\alpha p_e^\mu)\nonumber\\
	&&+c_8(i,j)(p^\mu p^\nu (p_\nu')^\alpha+p^\mu p^\alpha (p_\nu')^\nu+p^\nu p^\alpha (p_\nu')^\mu)+c_9(i,j)(p_e^\mu p_e^\nu p^\alpha+p_e^\mu p_e^\alpha p^\nu+p_e^\nu p_e^\alpha p^\mu)\nonumber\\
	&&+c_{10}(i,j)(p_e^\mu p_e^\nu (p_\nu')^\alpha+p_e^\mu p_e^\alpha (p_\nu')^\nu+p_e^\nu p_e^\alpha (p_\nu')^\mu)\nonumber\\
	&&+c_{11}(i,j)((p_\nu')^\mu (p_\nu')^\nu p^\alpha+(p_\nu')^\mu (p_\nu')^\alpha p^\nu+(p_\nu')^\nu (p_\nu')^\alpha p^\mu)\nonumber\\
	&&+c_{12}(i,j)((p_\nu')^\mu (p_\nu')^\nu p_e^\alpha+(p_\nu')^\mu (p_\nu')^\alpha p_e^\nu+(p_\nu')^\nu (p_\nu')^\alpha p_e^\mu)\nonumber\\
	&&+c_{13}(i,j)(p^\mu p_e^\nu (p_\nu')^\alpha+p^\mu p_e^\alpha (p_\nu')^\nu+p^\nu p_e^\alpha (p_\nu')^\mu+p^\nu p_e^\mu (p_\nu')^\alpha+p^\alpha p_e^\mu (p_\nu')^\nu+p^\alpha p_e^\nu (p_\nu')^\mu)~.\nonumber\\\label{eq:Imunualpha}
\end{eqnarray} 
Using the procedure outlined in the previous subsection, we can in principle deduce all the functions $c_1$--$c_{13}$ without inverting a $13\times 13$ matrix. However, as far as this work is concerned, we only need a fully-contracted integral $s_{e\mu}s_{n\nu}s_{p\alpha}I^{\mu\nu\alpha}_{i,j}$. Therefore, we may take a shortcut and start with the following doubly-contracted integral:
\begin{eqnarray}
	\tilde{I}^\mu_{i,j}&\equiv& s_{n\nu}s_{p\alpha}I^{\mu\nu\alpha}_{i,j}\nonumber\\
	&=&\tilde{c}_1(i,j)p^\mu+\tilde{c}_2(i,j)p_e^\mu+\tilde{c}_3(i,j)(p_\nu')^\mu+\tilde{c}_4(i,j)s_n^\mu+\tilde{c}_5(i,j)s_p^\mu~,\nonumber\\\label{eq:Itildemu}
\end{eqnarray}
where $\tilde{c}_4=c_2s_p\cdot p_e+c_3 s_p\cdot p_\nu'$ and $\tilde{c}_5=c_2 s_n\cdot p_e+c_3 s_n\cdot p_\nu'$ as following Eq.\eqref{eq:Imunualpha}. Now we play the same game: First, $c_2$ and $c_3$ can be obtained from $I^{zz\alpha}_{i,j}=-c_1(i,j)p^\alpha-c_2(i,j) p_e^\alpha-c_3(i,j)(p_\nu')^\alpha$, where $I^{zz\alpha}_{i,j}$ can be re-expressed in terms of $I^\mu$ using Eq.\eqref{eq:kz2}. Next, the three remaining functions $\tilde{c}_{1-3}$ can be deduced from the matrix equation:
\begin{equation}
	\left(\begin{array}{c}
		\tilde{c}_{1}(i,j)\\
		\tilde{c}_{2}(i,j)\\
		\tilde{c}_{3}(i,j)
	\end{array}\right)=M^{-1}\left(\begin{array}{c}
		s_{n\nu}s_{p\alpha}I_{i-1,j}^{\nu\alpha}\\
		s_{n\nu}s_{p\alpha}I_{i,j-1}^{\nu\alpha}-\tilde{c}_{4}(i,j)s_{n}\cdot p_{e}-\tilde{c}_{5}(i,j)s_{p}\cdot p_{e}\\
		(p_{\nu}^{\prime2}/2)s_{n\nu}s_{p\alpha}I_{i,j}^{\nu\alpha}-\tilde{c}_{4}(i,j)s_{n}\cdot p_{\nu}'-\tilde{c}_{5}(i,j)s_{p}\cdot p_{\nu}'
	\end{array}\right)~.
\end{equation}
Plugging the solution back into Eq.\eqref{eq:Itildemu} and contracting by $p_{e\mu}$ gives the desired fully-contracted integral:
\begin{eqnarray}
s_{e\mu}s_{n\nu}s_{p\alpha}I^{\mu\nu\alpha}_{i,j}&=&\mathcal{K}_1(i,j)s_n\cdot s_p s_e\cdot p+\mathcal{K}_2(i,j)s_n\cdot p_e s_p\cdot p_e s_e\cdot p+\mathcal{K}_3s_n\cdot p_\nu' s_p\cdot p_\nu' s_e\cdot p\nonumber\\
&&+\mathcal{K}_4(i,j)(s_n\cdot p_e s_p\cdot p_\nu'+s_n\cdot p_\nu' s_p\cdot p_e)s_e\cdot p+\mathcal{K}_5(i,j)s_n\cdot s_p s_e\cdot p_\nu'\nonumber\\
&&+\mathcal{K}_6(i,j)s_n\cdot p_e s_p\cdot p_e s_e\cdot p_\nu'+\mathcal{K}_7(i,j)s_n\cdot p_\nu' s_p\cdot p_\nu' s_e\cdot p_\nu'\nonumber\\
&&+\mathcal{K}_8(i,j)(s_n\cdot p_e s_p\cdot p_\nu'+s_n\cdot p_\nu' s_p\cdot p_e)s_e\cdot p_\nu'\nonumber\\
&&+\mathcal{K}_9(i,j)(s_p\cdot p_e s_n\cdot s_e+s_n\cdot p_e s_p\cdot s_e)\nonumber\\
&&+\mathcal{K}_{10}(i,j)(s_p\cdot p_\nu' s_n\cdot s_e+s_n\cdot p_\nu' s_p\cdot s_e)~.
\end{eqnarray}
The functions $\mathcal{K}_\eta$ can be decomposed as:
\begin{eqnarray}
	\mathcal{K}_\eta(i,j)&=&\mathcal{K}_{\eta a}I_{i,j}+\mathcal{K}_{\eta b}I_{i-1,j}+\mathcal{K}_{\eta c}I_{i,j-1}+\mathcal{K}_{\eta d}I_{i-2,j}+\mathcal{K}_{\eta e}I_{i,j-2}+\mathcal{K}_{\eta f}I_{i-1,j-1}\nonumber\\
	&&+\mathcal{K}_{\eta g}I_{i-3,j}+\mathcal{K}_{\eta h}I_{i,j-3}+\mathcal{K}_{\eta i}I_{i-2,j-1}+\mathcal{K}_{\eta j}I_{i-1,j-2}~,~\eta=1,...,10~,\label{eq:Kexpand}
\end{eqnarray}
where the analytic expressions of the coefficients $\{\mathcal{K}_{\eta a},...,\mathcal{K}_{\eta j}\}$ ($\eta=1,...,10$) can be found in \textbf{bremreg.nb}.

\section{Instructions for the supplemented \textit{Mathematica} notebooks}		

We provide two \textit{Mathematica} notebooks as supplementary materials: ``\textbf{recoil.nb}'' that evaluates analytically the recoil coefficients $\{r_{\eta a},...,r_{\eta d}\}$ in Eq.\eqref{eq:rcoeff}, and ``\textbf{bremreg.nb}'' that evaluates numerically the functions $\delta_{\eta}^\text{reg}(E_e,c')$ in Eq.\eqref{eq:gbremreg}. In this appendix we provide a brief explanation of their contents.

We used TRACER~\cite{Jamin:1991dp} to perform traces in the squared amplitude and impose on-shell conditions. The package can be downloaded from \url{https://library.wolfram.com/infocenter/MathSource/2987/} and is called in the first line of both notebooks.

\subsection{recoil.nb}

The notebook \textbf{recoil.nb} consists of three sections: 
\begin{itemize}
	\item The ``setup'' section defines the 3-body squared amplitude, imposes on-shell conditions and performs the $\epsilon$-expansion in Eq.\eqref{eq:epsilon}.
	\item The ``$\mathcal{O}(\epsilon^0)$'' section evaluates the zeroth-order correlation coefficients, i.e. $\mathcal{C}_0$ in Table~\ref{tab:RC1}-\ref{tab:RC2}.
	\item The ``$\mathcal{O}(\epsilon^1)$'' section evaluates the recoil coefficients $\{r_{\eta a},...,r_{\eta d}\}$ ($\eta=1,..,30$) in Table~\ref{tab:recoil1}-\ref{tab:recoil2}. 
\end{itemize}
Simply run the entire notebook to obtain the full result. For sceptical readers, important parts to check are the implementation of the on-shell conditions, the definition of $L_{\mu\nu}$, $H_{\mu\nu}$ and the $\epsilon$-expansion in the ``setup'' section, and the respective completeness check at the end of the $\mathcal{O}(\epsilon^0)$, $\mathcal{O}(\epsilon^1)$ sections.

\subsection{bremreg.nb}

The notebook \textbf{bremreg.nb} evaluates numerically the functions $\delta_\eta^\text{reg}(E_e,c')$ (denoted as $\delta\text{reg}\eta\text{[Ee,cp]}$ in the notebook) in Eq.\eqref{eq:gbremreg} ($\eta=1,...,36$). To use them, simply evaluate the entire notebook from the beginning to the ``regular bremsstrahlung'' section, which may take up to a few minutes. Below we briefly explain the purpose of each section:
\begin{itemize}
	\item The ``squared amplitude'' section constructs the ``regular'' bremsstrahlung squared amplitude $|\mathcal{M}_4|_\text{QM,reg}^2$ in Eq.\eqref{eq:M4reg}.
	\item The ``$(s\cdot k)^n$ replacement'' section applies the results in Appendix~\ref{sec:tensor} to transform the tensor integrals into scalar integrals. The key is to replace powers of $s\cdot k$ in $|\mathcal{M}_4|_\text{QM,reg}^2$ by powers of $p\cdot k$ and $p_e\cdot k$ which give the same result after integrating over $p_\nu$ and $k$. 
	\item The ``isolating different correlations'' section identifies the 36 structures in $g_\text{brem}^\text{reg}$ and check their completeness. It also performs the $p_\nu$, $k$-integral analytically by replacing $(p\cdot k)^{-i} (p_e\cdot k)^{-j}$ with the corresponding scalar integrals in Appendix~\ref{sec:scalar}.
	\item The ``functions and parameter'' section defines various inputs needed to evaluate the bremsstrahlung integral:
	\begin{itemize}
		\item The $3\times 3$ matrix  $M$ in Eq.\eqref{eq:Mmatrix} and its inverse, the coefficients $\{b_{\eta a},...,b_{\eta f}\}$ ($\eta=1,...,7$) in Eq.\eqref{eq:bexpand}, the coefficients $\{\mathcal{K}_{\eta a},...,\mathcal{K}_{\eta j}\}$ ($\eta=1,...,10$) in Eq.\eqref{eq:Kexpand}; 
		\item The scalar functions $I_{i,j}$ in Appendix~\ref{sec:scalar};
		\item The final integrand before performing the $\pnup$ integral (which takes most of the evaluation time).
	\end{itemize}
	\item The ``regular bremsstrahlung'' section defines various numerical inputs (fermion masses, $\lambda$, and the electron end-point energy), and obtain the functions $\delta_{\eta}^\text{reg}(E_e,c')$ by performing the $\pnup$-integral numerically. 
	\item Finally, the ``examples'' section demonstrates how one utilizes the functions $\delta_{\eta}^{\text{reg}}(E_e,c')$. In particular, the ``consistency test'' subsection checks the identities in Sec.\ref{sec:checkRC} numerically.  
\end{itemize}
A technical detail: The basis vectors $\hat{x}^\mu$ and $\hat{y}^\mu$ in Eqs.\eqref{eq:hatxmu}, \eqref{eq:hatymu} are, strictly speaking, undefined at $E_e=m_e$ or $c'=\pm 1$, which means the functions $\delta_{\eta}^{\text{reg}}(E_e,c')$ cannot be evaluated at these points, although their limits at $E_e\rightarrow m_e$ or $c'\rightarrow \pm 1$ are totally regular. 
\end{appendix}

\bibliography{ref}

\begin{thebibliography}{38}%
\makeatletter
\providecommand \@ifxundefined [1]{%
 \@ifx{#1\undefined}
}%
\providecommand \@ifnum [1]{%
 \ifnum #1\expandafter \@firstoftwo
 \else \expandafter \@secondoftwo
 \fi
}%
\providecommand \@ifx [1]{%
 \ifx #1\expandafter \@firstoftwo
 \else \expandafter \@secondoftwo
 \fi
}%
\providecommand \natexlab [1]{#1}%
\providecommand \enquote  [1]{``#1''}%
\providecommand \bibnamefont  [1]{#1}%
\providecommand \bibfnamefont [1]{#1}%
\providecommand \citenamefont [1]{#1}%
\providecommand \href@noop [0]{\@secondoftwo}%
\providecommand \href [0]{\begingroup \@sanitize@url \@href}%
\providecommand \@href[1]{\@@startlink{#1}\@@href}%
\providecommand \@@href[1]{\endgroup#1\@@endlink}%
\providecommand \@sanitize@url [0]{\catcode `\\12\catcode `\$12\catcode
  `\&12\catcode `\#12\catcode `\^12\catcode `\_12\catcode `\%12\relax}%
\providecommand \@@startlink[1]{}%
\providecommand \@@endlink[0]{}%
\providecommand \url  [0]{\begingroup\@sanitize@url \@url }%
\providecommand \@url [1]{\endgroup\@href {#1}{\urlprefix }}%
\providecommand \urlprefix  [0]{URL }%
\providecommand \Eprint [0]{\href }%
\providecommand \doibase [0]{https://doi.org/}%
\providecommand \selectlanguage [0]{\@gobble}%
\providecommand \bibinfo  [0]{\@secondoftwo}%
\providecommand \bibfield  [0]{\@secondoftwo}%
\providecommand \translation [1]{[#1]}%
\providecommand \BibitemOpen [0]{}%
\providecommand \bibitemStop [0]{}%
\providecommand \bibitemNoStop [0]{.\EOS\space}%
\providecommand \EOS [0]{\spacefactor3000\relax}%
\providecommand \BibitemShut  [1]{\csname bibitem#1\endcsname}%
\let\auto@bib@innerbib\@empty
\bibitem [{\citenamefont {Seng}(2024{\natexlab{a}})}]{Seng:2024fvi}%
  \BibitemOpen
  \bibfield  {author} {\bibinfo {author} {\bibfnamefont {C.-Y.}\ \bibnamefont
  {Seng}},\ }\bibfield  {title} {\bibinfo {title} {{Testing effective field
  theory with the most general neutron decay correlations}},\ }\href
  {https://doi.org/10.1103/PhysRevD.109.073007} {\bibfield  {journal} {\bibinfo
   {journal} {Phys. Rev. D}\ }\textbf {\bibinfo {volume} {109}},\ \bibinfo
  {pages} {073007} (\bibinfo {year} {2024}{\natexlab{a}})},\ \Eprint
  {https://arxiv.org/abs/2403.05714} {arXiv:2403.05714 [hep-ph]} \BibitemShut
  {NoStop}%
\bibitem [{\citenamefont {Beck}\ \emph {et~al.}(2020)\citenamefont {Beck} \emph
  {et~al.}}]{Beck:2019xye}%
  \BibitemOpen
  \bibfield  {author} {\bibinfo {author} {\bibfnamefont {M.}~\bibnamefont
  {Beck}} \emph {et~al.},\ }\bibfield  {title} {\bibinfo {title} {{Improved
  determination of the $\beta$-$\overline{\nu}_e$ angular correlation
  coefficient $a$ in free neutron decay with the $aSPECT$ spectrometer}},\
  }\href {https://doi.org/10.1103/PhysRevC.101.055506} {\bibfield  {journal}
  {\bibinfo  {journal} {Phys. Rev. C}\ }\textbf {\bibinfo {volume} {101}},\
  \bibinfo {pages} {055506} (\bibinfo {year} {2020})},\ \Eprint
  {https://arxiv.org/abs/1908.04785} {arXiv:1908.04785 [nucl-ex]} \BibitemShut
  {NoStop}%
\bibitem [{\citenamefont {Beck}\ \emph {et~al.}(2024)\citenamefont {Beck},
  \citenamefont {Heil}, \citenamefont {Schmidt}, \citenamefont {Bae\ss{}ler},
  \citenamefont {Gl\"uck}, \citenamefont {Konrad},\ and\ \citenamefont
  {Schmidt}}]{Beck:2023hnt}%
  \BibitemOpen
  \bibfield  {author} {\bibinfo {author} {\bibfnamefont {M.}~\bibnamefont
  {Beck}}, \bibinfo {author} {\bibfnamefont {W.}~\bibnamefont {Heil}}, \bibinfo
  {author} {\bibfnamefont {C.}~\bibnamefont {Schmidt}}, \bibinfo {author}
  {\bibfnamefont {S.}~\bibnamefont {Bae\ss{}ler}}, \bibinfo {author}
  {\bibfnamefont {F.}~\bibnamefont {Gl\"uck}}, \bibinfo {author} {\bibfnamefont
  {G.}~\bibnamefont {Konrad}},\ and\ \bibinfo {author} {\bibfnamefont
  {U.}~\bibnamefont {Schmidt}},\ }\bibfield  {title} {\bibinfo {title}
  {{Reanalysis of the
  \ensuremath{\beta}\ensuremath{-}\ensuremath{\nu}\textasciimacron{}e Angular
  Correlation Measurement from the aSPECT Experiment with New Constraints on
  Fierz Interference}},\ }\href
  {https://doi.org/10.1103/PhysRevLett.132.102501} {\bibfield  {journal}
  {\bibinfo  {journal} {Phys. Rev. Lett.}\ }\textbf {\bibinfo {volume} {132}},\
  \bibinfo {pages} {102501} (\bibinfo {year} {2024})},\ \Eprint
  {https://arxiv.org/abs/2308.16170} {arXiv:2308.16170 [nucl-ex]} \BibitemShut
  {NoStop}%
\bibitem [{\citenamefont {M\"arkisch}\ \emph {et~al.}(2019)\citenamefont
  {M\"arkisch} \emph {et~al.}}]{Markisch:2018ndu}%
  \BibitemOpen
  \bibfield  {author} {\bibinfo {author} {\bibfnamefont {B.}~\bibnamefont
  {M\"arkisch}} \emph {et~al.},\ }\bibfield  {title} {\bibinfo {title}
  {{Measurement of the Weak Axial-Vector Coupling Constant in the Decay of Free
  Neutrons Using a Pulsed Cold Neutron Beam}},\ }\href
  {https://doi.org/10.1103/PhysRevLett.122.242501} {\bibfield  {journal}
  {\bibinfo  {journal} {Phys. Rev. Lett.}\ }\textbf {\bibinfo {volume} {122}},\
  \bibinfo {pages} {242501} (\bibinfo {year} {2019})},\ \Eprint
  {https://arxiv.org/abs/1812.04666} {arXiv:1812.04666 [nucl-ex]} \BibitemShut
  {NoStop}%
\bibitem [{\citenamefont {Fermi}(1934)}]{Fermi:1934hr}%
  \BibitemOpen
  \bibfield  {author} {\bibinfo {author} {\bibfnamefont {E.}~\bibnamefont
  {Fermi}},\ }\bibfield  {title} {\bibinfo {title} {{An attempt of a theory of
  beta radiation. 1.}},\ }\href {https://doi.org/10.1007/BF01351864} {\bibfield
   {journal} {\bibinfo  {journal} {Z. Phys.}\ }\textbf {\bibinfo {volume}
  {88}},\ \bibinfo {pages} {161} (\bibinfo {year} {1934})}\BibitemShut
  {NoStop}%
\bibitem [{\citenamefont {Gl{\"u}ck}(2023)}]{Gluck:2022ogz}%
  \BibitemOpen
  \bibfield  {author} {\bibinfo {author} {\bibfnamefont {F.}~\bibnamefont
  {Gl{\"u}ck}},\ }\bibfield  {title} {\bibinfo {title} {{Radiative corrections
  to neutron and nuclear \ensuremath{\beta}-decays: a serious kinematics
  problem in the literature}},\ }\href
  {https://doi.org/10.1007/JHEP09(2023)188} {\bibfield  {journal} {\bibinfo
  {journal} {JHEP}\ }\textbf {\bibinfo {volume} {09}},\ \bibinfo {pages}
  {188}},\ \Eprint {https://arxiv.org/abs/2205.05042} {arXiv:2205.05042
  [hep-ph]} \BibitemShut {NoStop}%
\bibitem [{\citenamefont {Sirlin}(1967)}]{Sirlin:1967zza}%
  \BibitemOpen
  \bibfield  {author} {\bibinfo {author} {\bibfnamefont {A.}~\bibnamefont
  {Sirlin}},\ }\bibfield  {title} {\bibinfo {title} {{General Properties of the
  Electromagnetic Corrections to the Beta Decay of a Physical Nucleon}},\
  }\href {https://doi.org/10.1103/PhysRev.164.1767} {\bibfield  {journal}
  {\bibinfo  {journal} {Phys. Rev.}\ }\textbf {\bibinfo {volume} {164}},\
  \bibinfo {pages} {1767} (\bibinfo {year} {1967})}\BibitemShut {NoStop}%
\bibitem [{\citenamefont {Shann}(1971)}]{Shann:1971fz}%
  \BibitemOpen
  \bibfield  {author} {\bibinfo {author} {\bibfnamefont {R.~T.}\ \bibnamefont
  {Shann}},\ }\bibfield  {title} {\bibinfo {title} {{Electromagnetic effects in
  the decay of polarized neutrons}},\ }\href
  {https://doi.org/10.1007/BF02734566} {\bibfield  {journal} {\bibinfo
  {journal} {Nuovo Cim. A}\ }\textbf {\bibinfo {volume} {5}},\ \bibinfo {pages}
  {591} (\bibinfo {year} {1971})}\BibitemShut {NoStop}%
\bibitem [{\citenamefont {Garcia}\ and\ \citenamefont
  {Maya}(1978)}]{Garcia:1978bq}%
  \BibitemOpen
  \bibfield  {author} {\bibinfo {author} {\bibfnamefont {A.}~\bibnamefont
  {Garcia}}\ and\ \bibinfo {author} {\bibfnamefont {M.}~\bibnamefont {Maya}},\
  }\bibfield  {title} {\bibinfo {title} {{First Order Radiative Corrections to
  Asymmetry Coefficients in Neutron Decay}},\ }\href
  {https://doi.org/10.1103/PhysRevD.17.1376} {\bibfield  {journal} {\bibinfo
  {journal} {Phys. Rev. D}\ }\textbf {\bibinfo {volume} {17}},\ \bibinfo
  {pages} {1376} (\bibinfo {year} {1978})}\BibitemShut {NoStop}%
\bibitem [{\citenamefont {T\'oth}\ \emph {et~al.}(1986)\citenamefont {T\'oth},
  \citenamefont {Szego},\ and\ \citenamefont {Margaritisz}}]{Toth:1984er}%
  \BibitemOpen
  \bibfield  {author} {\bibinfo {author} {\bibfnamefont {K.}~\bibnamefont
  {T\'oth}}, \bibinfo {author} {\bibfnamefont {K.}~\bibnamefont {Szego}},\ and\
  \bibinfo {author} {\bibfnamefont {T.}~\bibnamefont {Margaritisz}},\
  }\bibfield  {title} {\bibinfo {title} {{Radiative Corrections for
  Semileptonic Decays of Hyperons: The 'Model Independent' Part}},\ }\href
  {https://doi.org/10.1103/PhysRevD.33.3306} {\bibfield  {journal} {\bibinfo
  {journal} {Phys. Rev. D}\ }\textbf {\bibinfo {volume} {33}},\ \bibinfo
  {pages} {3306} (\bibinfo {year} {1986})}\BibitemShut {NoStop}%
\bibitem [{\citenamefont {Gl{\"u}ck}(1993)}]{Gluck:1992tg}%
  \BibitemOpen
  \bibfield  {author} {\bibinfo {author} {\bibfnamefont {F.}~\bibnamefont
  {Gl{\"u}ck}},\ }\bibfield  {title} {\bibinfo {title} {{Measurable
  distributions of unpolarized neutron decay}},\ }\href
  {https://doi.org/10.1103/PhysRevD.47.2840} {\bibfield  {journal} {\bibinfo
  {journal} {Phys. Rev. D}\ }\textbf {\bibinfo {volume} {47}},\ \bibinfo
  {pages} {2840} (\bibinfo {year} {1993})}\BibitemShut {NoStop}%
\bibitem [{\citenamefont {Gl{\"u}ck}\ and\ \citenamefont
  {T\'oth}(1990)}]{Gluck:1989sf}%
  \BibitemOpen
  \bibfield  {author} {\bibinfo {author} {\bibfnamefont {F.}~\bibnamefont
  {Gl{\"u}ck}}\ and\ \bibinfo {author} {\bibfnamefont {K.}~\bibnamefont
  {T\'oth}},\ }\bibfield  {title} {\bibinfo {title} {{Order $\alpha$ Radiative
  Corrections for Semileptonic Decays of Unpolarized Baryons}},\ }\href
  {https://doi.org/10.1103/PhysRevD.54.1241} {\bibfield  {journal} {\bibinfo
  {journal} {Phys. Rev. D}\ }\textbf {\bibinfo {volume} {41}},\ \bibinfo
  {pages} {2160} (\bibinfo {year} {1990})},\ \bibinfo {note} {[Erratum:
  Phys.Rev.D 54, 1241 (1996)]}\BibitemShut {NoStop}%
\bibitem [{\citenamefont {Gl{\"u}ck}\ and\ \citenamefont
  {T\'oth}(1992)}]{Gluck:1992qy}%
  \BibitemOpen
  \bibfield  {author} {\bibinfo {author} {\bibfnamefont {F.}~\bibnamefont
  {Gl{\"u}ck}}\ and\ \bibinfo {author} {\bibfnamefont {K.}~\bibnamefont
  {T\'oth}},\ }\bibfield  {title} {\bibinfo {title} {{Order alpha radiative
  corrections for semileptonic decays of polarized baryons}},\ }\href
  {https://doi.org/10.1103/PhysRevD.46.2090} {\bibfield  {journal} {\bibinfo
  {journal} {Phys. Rev. D}\ }\textbf {\bibinfo {volume} {46}},\ \bibinfo
  {pages} {2090} (\bibinfo {year} {1992})}\BibitemShut {NoStop}%
\bibitem [{\citenamefont {Gl{\"u}ck}(1998)}]{Gluck:1998ogp}%
  \BibitemOpen
  \bibfield  {author} {\bibinfo {author} {\bibfnamefont {F.}~\bibnamefont
  {Gl{\"u}ck}},\ }\bibfield  {title} {\bibinfo {title} {{Electron spectra and
  electron-proton asymmetries in polarized neutron decay}},\ }\href
  {https://doi.org/10.1016/S0370-2693(98)00881-8} {\bibfield  {journal}
  {\bibinfo  {journal} {Phys. Lett. B}\ }\textbf {\bibinfo {volume} {436}},\
  \bibinfo {pages} {25} (\bibinfo {year} {1998})}\BibitemShut {NoStop}%
\bibitem [{\citenamefont {Gl{\"u}ck}(1997)}]{Gluck:1997km}%
  \BibitemOpen
  \bibfield  {author} {\bibinfo {author} {\bibfnamefont {F.}~\bibnamefont
  {Gl{\"u}ck}},\ }\bibfield  {title} {\bibinfo {title} {{Order-alpha radiative
  correction calculations for unoriented allowed nuclear, neutron and pion beta
  decays}},\ }\href {https://doi.org/10.1016/S0010-4655(96)00168-3} {\bibfield
  {journal} {\bibinfo  {journal} {Comput. Phys. Commun.}\ }\textbf {\bibinfo
  {volume} {101}},\ \bibinfo {pages} {223} (\bibinfo {year}
  {1997})}\BibitemShut {NoStop}%
\bibitem [{\citenamefont {Gl{\"u}ck}\ and\ \citenamefont
  {Joo}(1997)}]{Gluck:1994sw}%
  \BibitemOpen
  \bibfield  {author} {\bibinfo {author} {\bibfnamefont {F.}~\bibnamefont
  {Gl{\"u}ck}}\ and\ \bibinfo {author} {\bibfnamefont {I.}~\bibnamefont
  {Joo}},\ }\bibfield  {title} {\bibinfo {title} {{Hard photon Bremsstrahlung
  effects in hyperon semileptonic decays}},\ }\href
  {https://doi.org/10.1016/S0010-4655(97)00110-0} {\bibfield  {journal}
  {\bibinfo  {journal} {Comput. Phys. Commun.}\ }\textbf {\bibinfo {volume}
  {107}},\ \bibinfo {pages} {92} (\bibinfo {year} {1997})}\BibitemShut
  {NoStop}%
\bibitem [{\citenamefont {Seng}(2024{\natexlab{b}})}]{Seng:2023ynd}%
  \BibitemOpen
  \bibfield  {author} {\bibinfo {author} {\bibfnamefont {C.-Y.}\ \bibnamefont
  {Seng}},\ }\bibfield  {title} {\bibinfo {title} {{Pseudo-neutrino versus
  recoil formalism for 4-body phase space and applications to nuclear decay}},\
  }\href {https://doi.org/10.1103/PhysRevC.109.035501} {\bibfield  {journal}
  {\bibinfo  {journal} {Phys. Rev. C}\ }\textbf {\bibinfo {volume} {109}},\
  \bibinfo {pages} {035501} (\bibinfo {year} {2024}{\natexlab{b}})},\ \Eprint
  {https://arxiv.org/abs/2312.08630} {arXiv:2312.08630 [nucl-th]} \BibitemShut
  {NoStop}%
\bibitem [{\citenamefont {Ando}\ \emph {et~al.}(2004)\citenamefont {Ando},
  \citenamefont {Fearing}, \citenamefont {Gudkov}, \citenamefont {Kubodera},
  \citenamefont {Myhrer}, \citenamefont {Nakamura},\ and\ \citenamefont
  {Sato}}]{Ando:2004rk}%
  \BibitemOpen
  \bibfield  {author} {\bibinfo {author} {\bibfnamefont {S.}~\bibnamefont
  {Ando}}, \bibinfo {author} {\bibfnamefont {H.~W.}\ \bibnamefont {Fearing}},
  \bibinfo {author} {\bibfnamefont {V.~P.}\ \bibnamefont {Gudkov}}, \bibinfo
  {author} {\bibfnamefont {K.}~\bibnamefont {Kubodera}}, \bibinfo {author}
  {\bibfnamefont {F.}~\bibnamefont {Myhrer}}, \bibinfo {author} {\bibfnamefont
  {S.}~\bibnamefont {Nakamura}},\ and\ \bibinfo {author} {\bibfnamefont
  {T.}~\bibnamefont {Sato}},\ }\bibfield  {title} {\bibinfo {title} {{Neutron
  beta decay in effective field theory}},\ }\href
  {https://doi.org/10.1016/j.physletb.2004.06.037} {\bibfield  {journal}
  {\bibinfo  {journal} {Phys. Lett. B}\ }\textbf {\bibinfo {volume} {595}},\
  \bibinfo {pages} {250} (\bibinfo {year} {2004})},\ \Eprint
  {https://arxiv.org/abs/nucl-th/0402100} {arXiv:nucl-th/0402100} \BibitemShut
  {NoStop}%
\bibitem [{\citenamefont {Tishchenko}\ \emph {et~al.}(2013)\citenamefont
  {Tishchenko} \emph {et~al.}}]{MuLan:2012sih}%
  \BibitemOpen
  \bibfield  {author} {\bibinfo {author} {\bibfnamefont {V.}~\bibnamefont
  {Tishchenko}} \emph {et~al.} (\bibinfo {collaboration} {MuLan}),\ }\bibfield
  {title} {\bibinfo {title} {{Detailed Report of the MuLan Measurement of the
  Positive Muon Lifetime and Determination of the Fermi Constant}},\ }\href
  {https://doi.org/10.1103/PhysRevD.87.052003} {\bibfield  {journal} {\bibinfo
  {journal} {Phys. Rev. D}\ }\textbf {\bibinfo {volume} {87}},\ \bibinfo
  {pages} {052003} (\bibinfo {year} {2013})},\ \Eprint
  {https://arxiv.org/abs/1211.0960} {arXiv:1211.0960 [hep-ex]} \BibitemShut
  {NoStop}%
\bibitem [{\citenamefont {Cabibbo}(1963)}]{Cabibbo:1963yz}%
  \BibitemOpen
  \bibfield  {author} {\bibinfo {author} {\bibfnamefont {N.}~\bibnamefont
  {Cabibbo}},\ }\bibfield  {title} {\bibinfo {title} {{Unitary Symmetry and
  Leptonic Decays}},\ }\bibfield  {booktitle} {\emph {\bibinfo {booktitle}
  {{Meeting of the Italian School of Physics and Weak Interactions Bologna,
  Italy, April 26-28, 1984}}},\ }\href
  {https://doi.org/10.1103/PhysRevLett.10.531} {\bibfield  {journal} {\bibinfo
  {journal} {Phys. Rev. Lett.}\ }\textbf {\bibinfo {volume} {10}},\ \bibinfo
  {pages} {531} (\bibinfo {year} {1963})}\BibitemShut {NoStop}%
\bibitem [{\citenamefont {Kobayashi}\ and\ \citenamefont
  {Maskawa}(1973)}]{Kobayashi:1973fv}%
  \BibitemOpen
  \bibfield  {author} {\bibinfo {author} {\bibfnamefont {M.}~\bibnamefont
  {Kobayashi}}\ and\ \bibinfo {author} {\bibfnamefont {T.}~\bibnamefont
  {Maskawa}},\ }\bibfield  {title} {\bibinfo {title} {{CP Violation in the
  Renormalizable Theory of Weak Interaction}},\ }\href
  {https://doi.org/10.1143/PTP.49.652} {\bibfield  {journal} {\bibinfo
  {journal} {Prog. Theor. Phys.}\ }\textbf {\bibinfo {volume} {49}},\ \bibinfo
  {pages} {652} (\bibinfo {year} {1973})}\BibitemShut {NoStop}%
\bibitem [{\citenamefont {Seng}\ \emph {et~al.}(2023)\citenamefont {Seng},
  \citenamefont {Cirigliano}, \citenamefont {Feng}, \citenamefont {Gorchtein},
  \citenamefont {Jin},\ and\ \citenamefont {Miller}}]{Seng:2023jby}%
  \BibitemOpen
  \bibfield  {author} {\bibinfo {author} {\bibfnamefont {C.-Y.}\ \bibnamefont
  {Seng}}, \bibinfo {author} {\bibfnamefont {V.}~\bibnamefont {Cirigliano}},
  \bibinfo {author} {\bibfnamefont {X.}~\bibnamefont {Feng}}, \bibinfo {author}
  {\bibfnamefont {M.}~\bibnamefont {Gorchtein}}, \bibinfo {author}
  {\bibfnamefont {L.}~\bibnamefont {Jin}},\ and\ \bibinfo {author}
  {\bibfnamefont {G.~A.}\ \bibnamefont {Miller}},\ }\bibfield  {title}
  {\bibinfo {title} {{Quark mass difference effects in hadronic Fermi matrix
  elements from first principles}},\ }\href
  {https://doi.org/10.1016/j.physletb.2023.138259} {\bibfield  {journal}
  {\bibinfo  {journal} {Phys. Lett. B}\ }\textbf {\bibinfo {volume} {846}},\
  \bibinfo {pages} {138259} (\bibinfo {year} {2023})},\ \Eprint
  {https://arxiv.org/abs/2306.10199} {arXiv:2306.10199 [hep-ph]} \BibitemShut
  {NoStop}%
\bibitem [{\citenamefont {Seng}\ \emph {et~al.}(2018)\citenamefont {Seng},
  \citenamefont {Gorchtein}, \citenamefont {Patel},\ and\ \citenamefont
  {Ramsey-Musolf}}]{Seng:2018yzq}%
  \BibitemOpen
  \bibfield  {author} {\bibinfo {author} {\bibfnamefont {C.-Y.}\ \bibnamefont
  {Seng}}, \bibinfo {author} {\bibfnamefont {M.}~\bibnamefont {Gorchtein}},
  \bibinfo {author} {\bibfnamefont {H.~H.}\ \bibnamefont {Patel}},\ and\
  \bibinfo {author} {\bibfnamefont {M.~J.}\ \bibnamefont {Ramsey-Musolf}},\
  }\bibfield  {title} {\bibinfo {title} {{Reduced Hadronic Uncertainty in the
  Determination of $V_{ud}$}},\ }\href
  {https://doi.org/10.1103/PhysRevLett.121.241804} {\bibfield  {journal}
  {\bibinfo  {journal} {Phys. Rev. Lett.}\ }\textbf {\bibinfo {volume} {121}},\
  \bibinfo {pages} {241804} (\bibinfo {year} {2018})},\ \Eprint
  {https://arxiv.org/abs/1807.10197} {arXiv:1807.10197 [hep-ph]} \BibitemShut
  {NoStop}%
\bibitem [{\citenamefont {Seng}\ \emph {et~al.}(2019)\citenamefont {Seng},
  \citenamefont {Gorchtein},\ and\ \citenamefont
  {Ramsey-Musolf}}]{Seng:2018qru}%
  \BibitemOpen
  \bibfield  {author} {\bibinfo {author} {\bibfnamefont {C.~Y.}\ \bibnamefont
  {Seng}}, \bibinfo {author} {\bibfnamefont {M.}~\bibnamefont {Gorchtein}},\
  and\ \bibinfo {author} {\bibfnamefont {M.~J.}\ \bibnamefont
  {Ramsey-Musolf}},\ }\bibfield  {title} {\bibinfo {title} {{Dispersive
  evaluation of the inner radiative correction in neutron and nuclear $\beta$
  decay}},\ }\href {https://doi.org/10.1103/PhysRevD.100.013001} {\bibfield
  {journal} {\bibinfo  {journal} {Phys. Rev.}\ }\textbf {\bibinfo {volume}
  {D100}},\ \bibinfo {pages} {013001} (\bibinfo {year} {2019})},\ \Eprint
  {https://arxiv.org/abs/1812.03352} {arXiv:1812.03352 [nucl-th]} \BibitemShut
  {NoStop}%
\bibitem [{\citenamefont {Czarnecki}\ \emph {et~al.}(2019)\citenamefont
  {Czarnecki}, \citenamefont {Marciano},\ and\ \citenamefont
  {Sirlin}}]{Czarnecki:2019mwq}%
  \BibitemOpen
  \bibfield  {author} {\bibinfo {author} {\bibfnamefont {A.}~\bibnamefont
  {Czarnecki}}, \bibinfo {author} {\bibfnamefont {W.~J.}\ \bibnamefont
  {Marciano}},\ and\ \bibinfo {author} {\bibfnamefont {A.}~\bibnamefont
  {Sirlin}},\ }\bibfield  {title} {\bibinfo {title} {{Radiative Corrections to
  Neutron and Nuclear Beta Decays Revisited}},\ }\href
  {https://doi.org/10.1103/PhysRevD.100.073008} {\bibfield  {journal} {\bibinfo
   {journal} {Phys. Rev. D}\ }\textbf {\bibinfo {volume} {100}},\ \bibinfo
  {pages} {073008} (\bibinfo {year} {2019})},\ \Eprint
  {https://arxiv.org/abs/1907.06737} {arXiv:1907.06737 [hep-ph]} \BibitemShut
  {NoStop}%
\bibitem [{\citenamefont {Seng}\ \emph {et~al.}(2020)\citenamefont {Seng},
  \citenamefont {Feng}, \citenamefont {Gorchtein},\ and\ \citenamefont
  {Jin}}]{Seng:2020wjq}%
  \BibitemOpen
  \bibfield  {author} {\bibinfo {author} {\bibfnamefont {C.-Y.}\ \bibnamefont
  {Seng}}, \bibinfo {author} {\bibfnamefont {X.}~\bibnamefont {Feng}}, \bibinfo
  {author} {\bibfnamefont {M.}~\bibnamefont {Gorchtein}},\ and\ \bibinfo
  {author} {\bibfnamefont {L.-C.}\ \bibnamefont {Jin}},\ }\bibfield  {title}
  {\bibinfo {title} {{Joint lattice QCD--dispersion theory analysis confirms
  the quark-mixing top-row unitarity deficit}},\ }\href
  {https://doi.org/10.1103/PhysRevD.101.111301} {\bibfield  {journal} {\bibinfo
   {journal} {Phys. Rev. D}\ }\textbf {\bibinfo {volume} {101}},\ \bibinfo
  {pages} {111301} (\bibinfo {year} {2020})},\ \Eprint
  {https://arxiv.org/abs/2003.11264} {arXiv:2003.11264 [hep-ph]} \BibitemShut
  {NoStop}%
\bibitem [{\citenamefont {Hayen}(2021)}]{Hayen:2020cxh}%
  \BibitemOpen
  \bibfield  {author} {\bibinfo {author} {\bibfnamefont {L.}~\bibnamefont
  {Hayen}},\ }\bibfield  {title} {\bibinfo {title} {{Standard model
  $\mathcal{O}(\alpha)$ renormalization of $g_A$ and its impact on new physics
  searches}},\ }\href {https://doi.org/10.1103/PhysRevD.103.113001} {\bibfield
  {journal} {\bibinfo  {journal} {Phys. Rev. D}\ }\textbf {\bibinfo {volume}
  {103}},\ \bibinfo {pages} {113001} (\bibinfo {year} {2021})},\ \Eprint
  {https://arxiv.org/abs/2010.07262} {arXiv:2010.07262 [hep-ph]} \BibitemShut
  {NoStop}%
\bibitem [{\citenamefont {Cirigliano}\ \emph {et~al.}(2023)\citenamefont
  {Cirigliano}, \citenamefont {Dekens}, \citenamefont {Mereghetti},\ and\
  \citenamefont {Tomalak}}]{Cirigliano:2023fnz}%
  \BibitemOpen
  \bibfield  {author} {\bibinfo {author} {\bibfnamefont {V.}~\bibnamefont
  {Cirigliano}}, \bibinfo {author} {\bibfnamefont {W.}~\bibnamefont {Dekens}},
  \bibinfo {author} {\bibfnamefont {E.}~\bibnamefont {Mereghetti}},\ and\
  \bibinfo {author} {\bibfnamefont {O.}~\bibnamefont {Tomalak}},\ }\bibfield
  {title} {\bibinfo {title} {{Effective field theory for radiative corrections
  to charged-current processes: Vector coupling}},\ }\href
  {https://doi.org/10.1103/PhysRevD.108.053003} {\bibfield  {journal} {\bibinfo
   {journal} {Phys. Rev. D}\ }\textbf {\bibinfo {volume} {108}},\ \bibinfo
  {pages} {053003} (\bibinfo {year} {2023})},\ \Eprint
  {https://arxiv.org/abs/2306.03138} {arXiv:2306.03138 [hep-ph]} \BibitemShut
  {NoStop}%
\bibitem [{\citenamefont {Gorchtein}\ and\ \citenamefont
  {Seng}(2021)}]{Gorchtein:2021fce}%
  \BibitemOpen
  \bibfield  {author} {\bibinfo {author} {\bibfnamefont {M.}~\bibnamefont
  {Gorchtein}}\ and\ \bibinfo {author} {\bibfnamefont {C.-Y.}\ \bibnamefont
  {Seng}},\ }\bibfield  {title} {\bibinfo {title} {{Dispersion relation
  analysis of the radiative corrections to g$_{A}$ in the neutron
  \ensuremath{\beta}-decay}},\ }\href {https://doi.org/10.1007/JHEP10(2021)053}
  {\bibfield  {journal} {\bibinfo  {journal} {JHEP}\ }\textbf {\bibinfo
  {volume} {10}},\ \bibinfo {pages} {053}},\ \Eprint
  {https://arxiv.org/abs/2106.09185} {arXiv:2106.09185 [hep-ph]} \BibitemShut
  {NoStop}%
\bibitem [{\citenamefont {Cirigliano}\ \emph {et~al.}(2022)\citenamefont
  {Cirigliano}, \citenamefont {de~Vries}, \citenamefont {Hayen}, \citenamefont
  {Mereghetti},\ and\ \citenamefont {Walker-Loud}}]{Cirigliano:2022hob}%
  \BibitemOpen
  \bibfield  {author} {\bibinfo {author} {\bibfnamefont {V.}~\bibnamefont
  {Cirigliano}}, \bibinfo {author} {\bibfnamefont {J.}~\bibnamefont
  {de~Vries}}, \bibinfo {author} {\bibfnamefont {L.}~\bibnamefont {Hayen}},
  \bibinfo {author} {\bibfnamefont {E.}~\bibnamefont {Mereghetti}},\ and\
  \bibinfo {author} {\bibfnamefont {A.}~\bibnamefont {Walker-Loud}},\
  }\bibfield  {title} {\bibinfo {title} {{Pion-Induced Radiative Corrections to
  Neutron \ensuremath{\beta} Decay}},\ }\href
  {https://doi.org/10.1103/PhysRevLett.129.121801} {\bibfield  {journal}
  {\bibinfo  {journal} {Phys. Rev. Lett.}\ }\textbf {\bibinfo {volume} {129}},\
  \bibinfo {pages} {121801} (\bibinfo {year} {2022})},\ \Eprint
  {https://arxiv.org/abs/2202.10439} {arXiv:2202.10439 [nucl-th]} \BibitemShut
  {NoStop}%
\bibitem [{\citenamefont {Seng}(2024{\natexlab{c}})}]{Seng:2024ker}%
  \BibitemOpen
  \bibfield  {author} {\bibinfo {author} {\bibfnamefont {C.-Y.}\ \bibnamefont
  {Seng}},\ }\bibfield  {title} {\bibinfo {title} {{Hybrid analysis of
  radiative corrections to neutron decay with current algebra and effective
  field theory}},\ }\href@noop {} {\  (\bibinfo {year} {2024}{\natexlab{c}})},\
  \Eprint {https://arxiv.org/abs/2403.08976} {arXiv:2403.08976 [hep-ph]}
  \BibitemShut {NoStop}%
\bibitem [{\citenamefont {Gudkov}\ \emph {et~al.}(2006)\citenamefont {Gudkov},
  \citenamefont {Greene},\ and\ \citenamefont {Calarco}}]{Gudkov:2005bu}%
  \BibitemOpen
  \bibfield  {author} {\bibinfo {author} {\bibfnamefont {V.~P.}\ \bibnamefont
  {Gudkov}}, \bibinfo {author} {\bibfnamefont {G.~L.}\ \bibnamefont {Greene}},\
  and\ \bibinfo {author} {\bibfnamefont {J.~R.}\ \bibnamefont {Calarco}},\
  }\bibfield  {title} {\bibinfo {title} {{General classification and analysis
  of neutron beta-decay experiments}},\ }\href
  {https://doi.org/10.1103/PhysRevC.73.035501} {\bibfield  {journal} {\bibinfo
  {journal} {Phys. Rev. C}\ }\textbf {\bibinfo {volume} {73}},\ \bibinfo
  {pages} {035501} (\bibinfo {year} {2006})},\ \Eprint
  {https://arxiv.org/abs/nucl-th/0510012} {arXiv:nucl-th/0510012} \BibitemShut
  {NoStop}%
\bibitem [{\citenamefont {Bhattacharya}\ \emph {et~al.}(2012)\citenamefont
  {Bhattacharya}, \citenamefont {Cirigliano}, \citenamefont {Cohen},
  \citenamefont {Filipuzzi}, \citenamefont {Gonzalez-Alonso}, \citenamefont
  {Graesser}, \citenamefont {Gupta},\ and\ \citenamefont
  {Lin}}]{Bhattacharya:2011qm}%
  \BibitemOpen
  \bibfield  {author} {\bibinfo {author} {\bibfnamefont {T.}~\bibnamefont
  {Bhattacharya}}, \bibinfo {author} {\bibfnamefont {V.}~\bibnamefont
  {Cirigliano}}, \bibinfo {author} {\bibfnamefont {S.~D.}\ \bibnamefont
  {Cohen}}, \bibinfo {author} {\bibfnamefont {A.}~\bibnamefont {Filipuzzi}},
  \bibinfo {author} {\bibfnamefont {M.}~\bibnamefont {Gonzalez-Alonso}},
  \bibinfo {author} {\bibfnamefont {M.~L.}\ \bibnamefont {Graesser}}, \bibinfo
  {author} {\bibfnamefont {R.}~\bibnamefont {Gupta}},\ and\ \bibinfo {author}
  {\bibfnamefont {H.-W.}\ \bibnamefont {Lin}},\ }\bibfield  {title} {\bibinfo
  {title} {{Probing Novel Scalar and Tensor Interactions from (Ultra)Cold
  Neutrons to the LHC}},\ }\href {https://doi.org/10.1103/PhysRevD.85.054512}
  {\bibfield  {journal} {\bibinfo  {journal} {Phys. Rev. D}\ }\textbf {\bibinfo
  {volume} {85}},\ \bibinfo {pages} {054512} (\bibinfo {year} {2012})},\
  \Eprint {https://arxiv.org/abs/1110.6448} {arXiv:1110.6448 [hep-ph]}
  \BibitemShut {NoStop}%
\bibitem [{\citenamefont {Ivanov}\ \emph {et~al.}(2019)\citenamefont {Ivanov},
  \citenamefont {H\"ollwieser}, \citenamefont {Troitskaya}, \citenamefont
  {Wellenzohn},\ and\ \citenamefont {Berdnikov}}]{Ivanov:2018yir}%
  \BibitemOpen
  \bibfield  {author} {\bibinfo {author} {\bibfnamefont {A.~N.}\ \bibnamefont
  {Ivanov}}, \bibinfo {author} {\bibfnamefont {R.}~\bibnamefont
  {H\"ollwieser}}, \bibinfo {author} {\bibfnamefont {N.~I.}\ \bibnamefont
  {Troitskaya}}, \bibinfo {author} {\bibfnamefont {M.}~\bibnamefont
  {Wellenzohn}},\ and\ \bibinfo {author} {\bibfnamefont {Y.~A.}\ \bibnamefont
  {Berdnikov}},\ }\bibfield  {title} {\bibinfo {title} {{Tests of the standard
  model in neutron beta decay with polarized electrons and unpolarized neutrons
  and protons}},\ }\href {https://doi.org/10.1103/PhysRevD.104.059902}
  {\bibfield  {journal} {\bibinfo  {journal} {Phys. Rev. D}\ }\textbf {\bibinfo
  {volume} {99}},\ \bibinfo {pages} {053004} (\bibinfo {year} {2019})},\
  \bibinfo {note} {[Erratum: Phys.Rev.D 104, 059902 (2021)]},\ \Eprint
  {https://arxiv.org/abs/1811.04853} {arXiv:1811.04853 [hep-ph]} \BibitemShut
  {NoStop}%
\bibitem [{\citenamefont {Ivanov}\ \emph {et~al.}(2017)\citenamefont {Ivanov},
  \citenamefont {H\"ollwieser}, \citenamefont {Troitskaya}, \citenamefont
  {Wellenzohn},\ and\ \citenamefont {Berdnikov}}]{Ivanov:2017mnz}%
  \BibitemOpen
  \bibfield  {author} {\bibinfo {author} {\bibfnamefont {A.~N.}\ \bibnamefont
  {Ivanov}}, \bibinfo {author} {\bibfnamefont {R.}~\bibnamefont
  {H\"ollwieser}}, \bibinfo {author} {\bibfnamefont {N.~I.}\ \bibnamefont
  {Troitskaya}}, \bibinfo {author} {\bibfnamefont {M.}~\bibnamefont
  {Wellenzohn}},\ and\ \bibinfo {author} {\bibfnamefont {Y.~A.}\ \bibnamefont
  {Berdnikov}},\ }\bibfield  {title} {\bibinfo {title} {{Precision analysis of
  electron energy spectrum and angular distribution of neutron
  \ensuremath{\beta}\ensuremath{-} decay with polarized neutron and
  electron}},\ }\href {https://doi.org/10.1103/PhysRevC.104.069901} {\bibfield
  {journal} {\bibinfo  {journal} {Phys. Rev. C}\ }\textbf {\bibinfo {volume}
  {95}},\ \bibinfo {pages} {055502} (\bibinfo {year} {2017})},\ \bibinfo {note}
  {[Erratum: Phys.Rev.C 104, 069901 (2021)]},\ \Eprint
  {https://arxiv.org/abs/1705.07330} {arXiv:1705.07330 [hep-ph]} \BibitemShut
  {NoStop}%
\bibitem [{\citenamefont {Ivanov}\ \emph {et~al.}(2018)\citenamefont {Ivanov},
  \citenamefont {H\"ollwieser}, \citenamefont {Troitskaya}, \citenamefont
  {Wellenzohn},\ and\ \citenamefont {Berdnikov}}]{Ivanov:2018vmz}%
  \BibitemOpen
  \bibfield  {author} {\bibinfo {author} {\bibfnamefont {A.~N.}\ \bibnamefont
  {Ivanov}}, \bibinfo {author} {\bibfnamefont {R.}~\bibnamefont
  {H\"ollwieser}}, \bibinfo {author} {\bibfnamefont {N.~I.}\ \bibnamefont
  {Troitskaya}}, \bibinfo {author} {\bibfnamefont {M.}~\bibnamefont
  {Wellenzohn}},\ and\ \bibinfo {author} {\bibfnamefont {Y.~A.}\ \bibnamefont
  {Berdnikov}},\ }\bibfield  {title} {\bibinfo {title} {{Tests of the standard
  model in neutron $\beta$ decay with a polarized neutron and electron and an
  unpolarized proton}},\ }\href {https://doi.org/10.1103/PhysRevC.98.035503}
  {\bibfield  {journal} {\bibinfo  {journal} {Phys. Rev. C}\ }\textbf {\bibinfo
  {volume} {98}},\ \bibinfo {pages} {035503} (\bibinfo {year} {2018})},\
  \Eprint {https://arxiv.org/abs/1805.03880} {arXiv:1805.03880 [hep-ph]}
  \BibitemShut {NoStop}%
\bibitem [{\citenamefont {Ginsberg}(1967)}]{Ginsberg:1969jh}%
  \BibitemOpen
  \bibfield  {author} {\bibinfo {author} {\bibfnamefont {E.~S.}\ \bibnamefont
  {Ginsberg}},\ }\bibfield  {title} {\bibinfo {title} {{Radiative corrections
  to the k-l-3 +- dalitz plot}},\ }\href
  {https://doi.org/10.1103/PhysRev.162.1570} {\bibfield  {journal} {\bibinfo
  {journal} {Phys. Rev.}\ }\textbf {\bibinfo {volume} {162}},\ \bibinfo {pages}
  {1570} (\bibinfo {year} {1967})},\ \bibinfo {note} {[Erratum: Phys.Rev. 187,
  2280 (1969)]}\BibitemShut {NoStop}%
\bibitem [{\citenamefont {Jamin}\ and\ \citenamefont
  {Lautenbacher}(1993)}]{Jamin:1991dp}%
  \BibitemOpen
  \bibfield  {author} {\bibinfo {author} {\bibfnamefont {M.}~\bibnamefont
  {Jamin}}\ and\ \bibinfo {author} {\bibfnamefont {M.~E.}\ \bibnamefont
  {Lautenbacher}},\ }\bibfield  {title} {\bibinfo {title} {{TRACER: Version
  1.1: A Mathematica package for gamma algebra in arbitrary dimensions}},\
  }\href {https://doi.org/10.1016/0010-4655(93)90097-V} {\bibfield  {journal}
  {\bibinfo  {journal} {Comput. Phys. Commun.}\ }\textbf {\bibinfo {volume}
  {74}},\ \bibinfo {pages} {265} (\bibinfo {year} {1993})}\BibitemShut
  {NoStop}%
\end{thebibliography}%

\end{document}